\documentclass[journal=jacsat,manuscript=article]{achemso}

\usepackage[version=3]{mhchem} 
\usepackage{tikz}
\newcommand*\circled[1]{\tikz[baseline=(char.base)]{
            \node[shape=circle,draw,inner sep=0.5pt] (char) {#1};}}
\newcommand*\squared[1]{\tikz[baseline=(char.base)]{
            \node[shape=rectangle, draw, inner sep=2pt, minimum width=0.5em, rounded corners = 2] (char) {#1};}}

\newcommand{\beginsupplement}{%
        \setcounter{table}{0}
        \renewcommand{\thetable}{S\arabic{table}}%
        \setcounter{figure}{0}
        \renewcommand{\thefigure}{S\arabic{figure}}%
     }

\author{Ashutosh Patri}
\email{ashutosh.patri@polymtl.ca}
\affiliation[Polytechnique Montr\'eal]
{Department of Electrical Engineering, Polytechnique Montr\'eal, Montr\'eal, Canada}
\author{K\'{e}vin G. Cogn\'{e}e}
\affiliation{Center for Discovery and Innovation, City College of New York, New York, USA}
\affiliation{Department of Engineering Physics, Polytechnique Montr\'eal, Montr\'eal, Canada}
\author{Louis Haeberl\'{e}}
\affiliation[Polytechnique Montr\'eal]
{Department of Engineering Physics, Polytechnique Montr\'eal, Montr\'eal, Canada}
\author{Vinod Menon}
\affiliation{Center for Discovery and Innovation, City College of New York, New York, USA}
\author{Christophe Caloz}
\affiliation[ KU Leuven]
{Department of Electrical Engineering, KU Leuven, Leuven, Belgium}
\author{St\'{e}phane K\'{e}na-Cohen}
\email{s.kena-cohen@polymtl.ca}
\affiliation[Polytechnique Montr\'eal]
{Department of Engineering Physics, Polytechnique Montr\'eal, Montr\'eal, Canada}

\title {Photonic Gap Antennas Based on \\High Index-Contrast Slot-Waveguides}

\begin{document}
\begin{tocentry}
\includegraphics[width=1\textwidth]{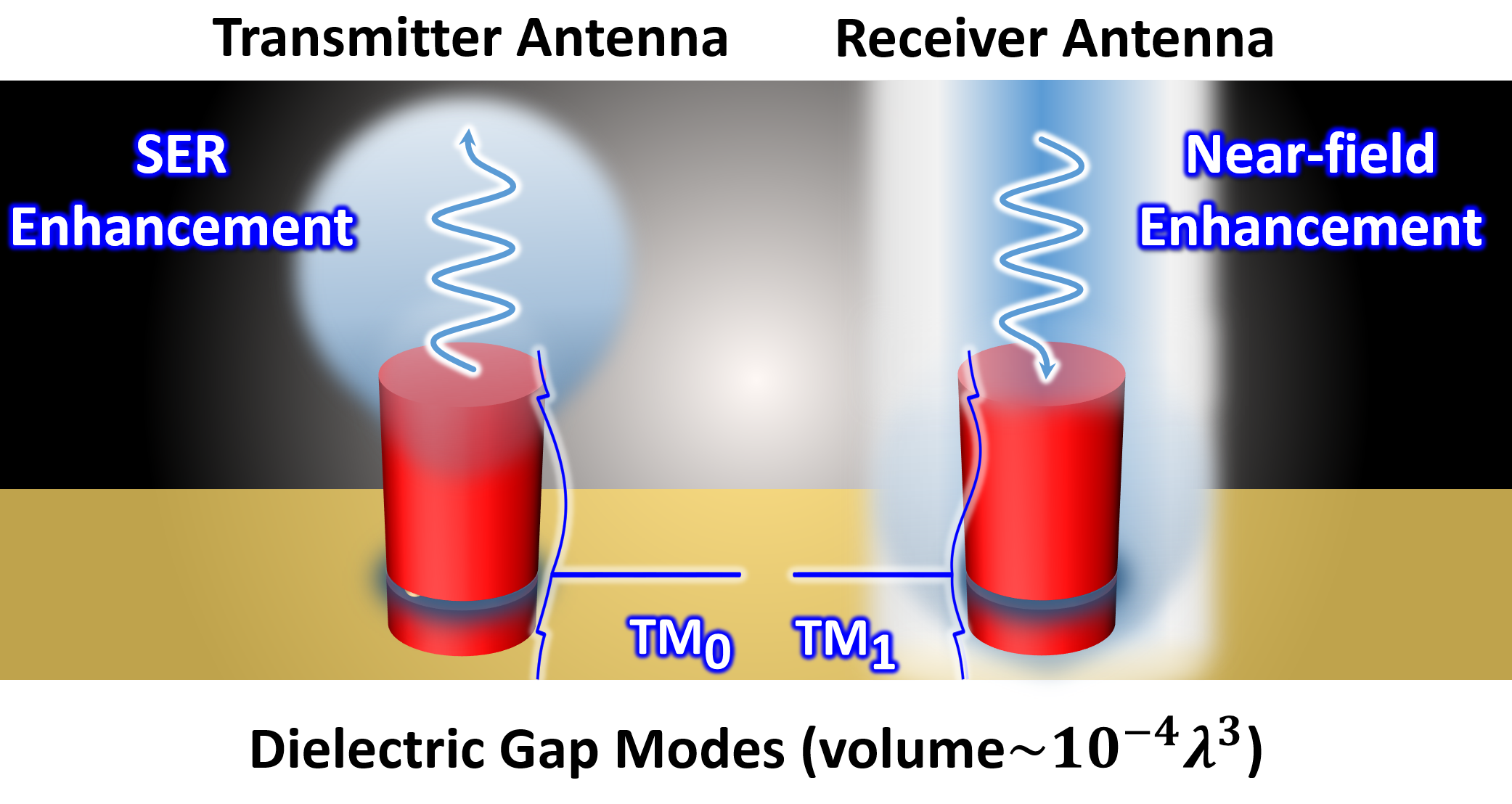}
\end{tocentry}

\begin{abstract}

Optical antennas made of low-loss dielectrics have several advantages over plasmonic antennas, including high radiative quantum efficiency, negligible heating and excellent photostability. However, due to weak spatial confinement, conventional dielectric antennas fail to offer light-matter interaction strengths on par with those of plasmonic antennas. We propose here an all-dielectric antenna configuration that can support strongly confined modes ($V\sim10^{-4}\lambda_{0}^3$) while maintaining unity antenna quantum efficiency. This configuration consists of a high-index pillar structure with a transverse gap that is filled with a low-index material, where the contrast of indices induces a strong enhancement of the electric field perpendicular to the gap. We provide a detailed explanation of the operation principle of such Photonic Gap Antennas (PGAs) based on the dispersion relation of symmetric and asymmetric horizontal slot-waveguides. To discuss the properties of PGAs, we consider silicon pillars with air or CYTOP as the gap-material. We show by full-wave simulations that PGAs with an emitter embedded in the gap can enhance the spontaneous emission rate by a factor of $\sim$1000 for air gaps and $\sim$400 for CYTOP gaps over a spectral bandwidth of $\Delta\lambda\approx300$ nm at $\lambda=1.25$ \textmu m. Furthermore, the PGAs can be designed to provide unidirectional out-of-plane radiation across a substantial portion of their spectral bandwidth. This is achieved by setting the position of the gap at an optimized off-centered position of the pillar so as to properly break the vertical symmetry of the structure. We also demonstrate that, when acting as receivers, PGAs can lead to a near-field intensity enhancement by a factor of $\sim$3000 for air gaps and $\sim$1200 for CYTOP gaps.

\end{abstract}

\section{Keywords} optical antennas, slot waveguides, dielectric nanoantennas, spontaneous emission rate, near-field enhancement, light-matter interaction

\section{Introduction}

The development of optical antennas has progressed tremendously over the past two decades. Similar to their long-wavelength (radio and microwave) counterparts, optical antennas convert far-field electromagnetic radiation into localized near-field components and vice versa. While transmission in long-wavelength antennas is driven by alternating electrical currents, optical antennas are typically excited by nanoscale emitters such as atoms, molecules or quantum dots. Therefore, an efficient extraction of electromagnetic radiation requires a strong localization of the near-field energy. To achieve this, the vast majority of theoretical and experimental work has focused on the use of sub-wavelength metallic antennas~\cite{novotny2011antennas}. Metals at visible and infrared frequencies support surface plasmons that allow for deep sub-wavelength ($\sim\lambda/10$) localization of oscillating electric fields. This is obtained by converting the electric field energy---the source of capacitance---into kinetic energy of free-electrons---the source of kinetic inductance~\cite{khurgin2012reflecting}. This contrasts with the case of long-wavelength and dielectric antennas, where energy oscillates mostly between the electric and magnetic fields. The ability of plasmonic antennas to manipulate or enhance the emission of nearby emitters has been found useful both for light-emitting devices~\cite{tsakmakidis2016large} and state-of-the-art single-photon sources~\cite{koenderink2017single}. Moreover, their ability to concentrate light has found important applications, for instance in sensing, nonlinear optics, integrated photonics and imaging~\cite{grober1997optical, bharadwaj2009optical, kauranen2012nonlinear, biagioni2012nanoantennas, agio2013optical}.

An important drawback of plasmonic antennas is the presence of ohmic losses due to various scattering processes that occur within the electron gas. This can lead to considerable heat generation, resulting in melting or irreversible structural alteration of the antenna and thermochemical destruction of the nearby matter~\cite{kuhlicke2013situ, caldarola2015non, mahmoudi2014variation,alessandri2016enhanced}. Moreover, losses within the metal implicitly limit the quantum efficiency of emitters, a phenomenon that severely hampers their utilization in applications where efficiency (loss) is important~\cite{barnes2002solid}. Although inefficient emitters can see their internal quantum efficiency increase via coupling to plasmonic antennas---because the antenna quantum efficiency can exceed that of the emitter---emitters that are \emph{a priori} efficient inevitably see their quantum efficiency lowered.

A strategy to avoid the ohmic losses in plasmonic antennas has been to use instead nanoantennas made of high refractive index dielectrics. Silicon nanospheres, for example, support strong Mie resonances in the visible and near-infrared~\cite{kuznetsov2016optically, bouchet2016enhancement}. However, since light in a dielectric material is bound by the diffraction limit, the volumetric modes (or bulk modes) of such antennas tend to suffer from very weak spatial confinement as compared to the surface modes of their metallic counterparts~\cite{bozhevolnyi2016fundamental}. In addition, bulk modes do not readily allow for the placement of emitters or analytes at the position of maximum near-field intensity~\cite{rutckaia2017quantum}. Recently, the use of multi-element all-dielectric structures such as dimers~\cite{albella2013low, regmi2016all} or oligomers~\cite{rocco2020giant}, where the dipole modes of individual elements hybridize, have been demonstrated to provide strong field confinement in the inter-element spacing region. Single-element gapped-structures that can support nonradiating anapole modes~\cite{yang2018anapole, mignuzzi2019nanoscale} with high quality factors have also been proposed to overcome these drawbacks. Nevertheless, the achievable electric-field confinement in such designs has remained limited by the lateral size of the nanoscale voids incorporated in the structure, whose resolution is dictated by the available nanolithography technology. Finally, for many applications, out-of-plane unidirectional radiation is desirable. Symmetric structures~\cite{yang2018anapole, mignuzzi2019nanoscale} inherently possess non-directional radiation patterns. Although breaking this symmetry to achieve directionality is possible with multiple lithography steps, this places stringent fabrication constraints~\cite{albella2015switchable}.

In this work, we propose dielectric antennas, which we refer to as Photonic Gap Antennas (PGAs) based on Fabry-Perot type resonances of slot-waveguide modes~\cite{almeida2004guiding}. We exploit the high field confinement capability of such dielectric gap modes~\cite{galli2006direct, galli2006strong, jun2009broadband, kolchin2015high, sakib2020design, robinson2005ultrasmall, choi2017self, hu2018experimental} using a simple multilayer pillar architecture, as shown in Fig.~\ref{fig:Dispersion}(a), and demonstrate spontaneous emission rate (SER) enhancements \textgreater1000 for an emitter embedded within the gap. We design silicon ($n=3.53$) nanopillar-based PGAs consisting of a gap layer of air ($n=1$) or CYTOP ($n=1.33$). The adoption of horizontal gap layers in PGAs, realizable by simple deposition processes, avoids the fabrication constraints of lithography-based techniques~\cite{lee2010silicon, sun2007horizontal, miyazaki2006squeezing}. We study the propagating eigenmodes of both vertically symmetric and asymmetric horizontal slot-waveguides and show that the corresponding resonant modes of finite-length structures can interfere with each other to realize unidirectional out-of-plane radiation. This directionality spans over a substantial portion of the emission enhancement bandwidth of $\Delta\lambda\approx300$~nm. We also discuss the scattering properties of PGAs, in the receiving regime, for an incident plane wave, and demonstrate that a field intensity enhancement as high as $\sim$3000 can be achieved. In a forthcoming paper, we consider the use of gaps with a vanishingly small refractive index ($n\approx0$) to explore the limit of spatial confinement using the gap modes. In that case, the nanopillar modes strongly hybridize with the epsilon-near-zero Berreman mode of the gap leading to extremely efficient light-matter interaction and intrinsic unidirectional radiation.

\section{PGA Design Principle}

\subsection{Gap Modes}

\begin{figure}[b]
\centering 
\includegraphics[width=0.85\textwidth]{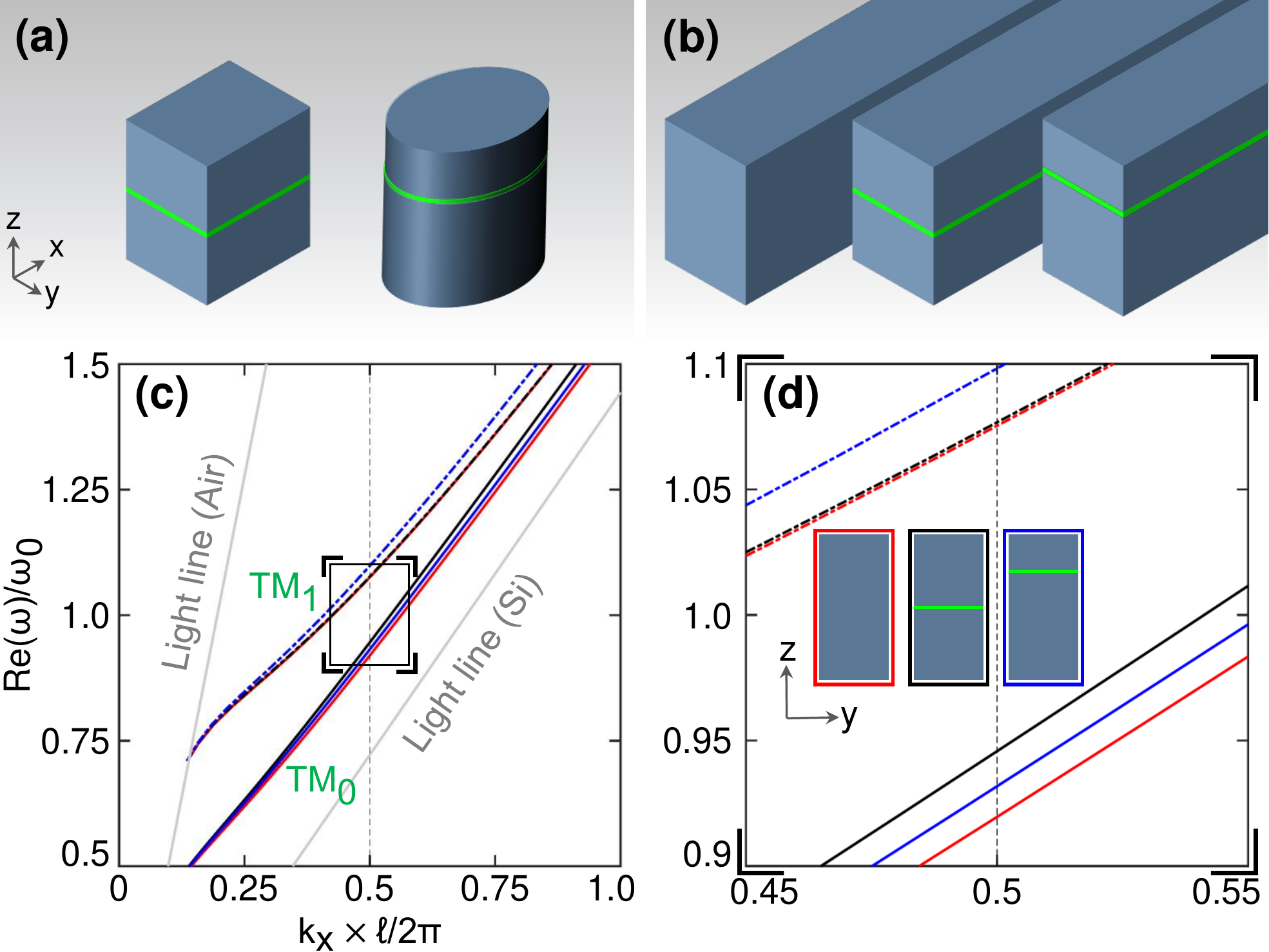}
\caption{Photonic Gap Antennas (PGAs) and dispersion relation of their infinite-length waveguide counterparts. (a) Perspective view of different PGA structures: a symmetric rectangular PGA (left) and an asymmetric elliptical PGA (right). (b) Perspective view of an infinite-length (along $x$) conventional silicon waveguide (left), a symmetric slot-waveguide (center), and asymmetric slot-waveguide (right). (c) Dispersion relation for the two lowest quasi-TM eigenmodes (electric field along the $z$-axis) showing the normalized real angular frequency, Re$(\omega)/\omega_0 ~\text{, where}~ \omega_0=2\pi (235)$~THz, as a function of the normalized propagation constant, $k_x\ell/2\pi ~\text{, where}~ \ell=250$~nm. These relations are plotted for the the conventional silicon waveguide (red), the symmetric slot-waveguide (black), and the asymmetric slot-waveguide (blue). The corresponding cross-sections are shown as insets in (d). The waveguides are composed of silicon, with an air gap, with a gap thickness of 2 nm. The height (including the gap) and width of the waveguide cross-sections are 582 nm (along $z$) and 240 nm (along $y$), respectively. The ratio ($t_1/t_2$) of the below-gap thickness ($t_1$) to the above-gap-thickness ($t_2$) of silicon are 1 and 3, respectively, for the symmetric and the asymmetric slot-waveguides. The solid grey lines show the light lines for bulk air and silicon. The vertical dashed line at $k_x\ell/2\pi=0.5$ intersects the dispersion curves of waveguides at their respective resonant frequencies for a finite-length ($\ell=250$~nm) of the structure along the $x$-axis. (d) Zoomed-in portion of the dispersion relation in (c), identified by a rectangle, highlighting changes due to the incorporation of the gap layer.}\label{fig:Dispersion}
\end{figure}

\begin{figure}[b]
\centering 
\includegraphics[width=0.75\textwidth]{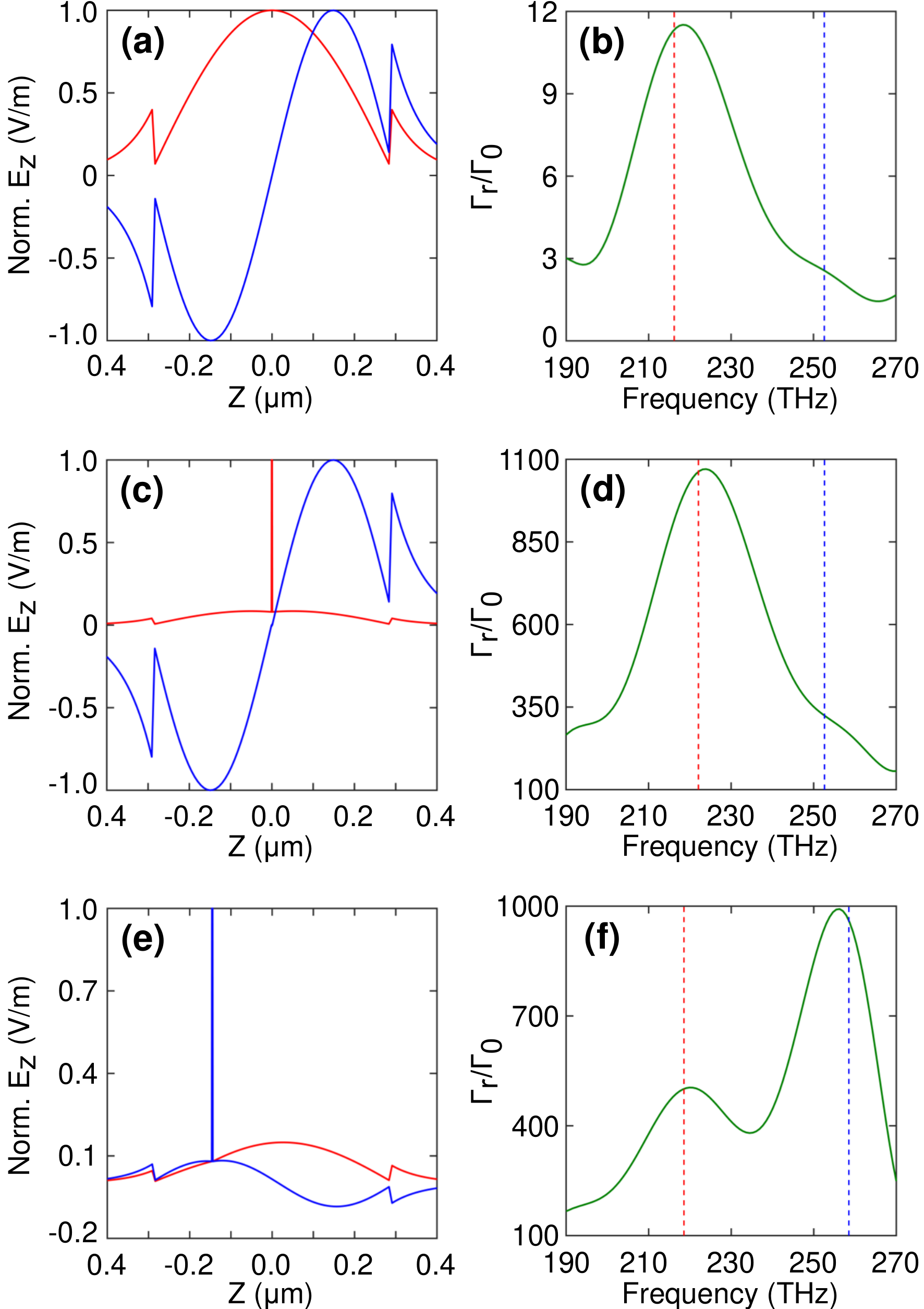}
\caption{Normalized $E_z$ of the eigenmode profiles for the infinite-length waveguides in Fig.~\ref{fig:Dispersion}(b) and SER enhancement factors for the corresponding PGAs (finite-length waveguide sections). The $E_z$ component for the $\text{TM}_0$ (red) and $\text{TM}_1$ (blue) modes are shown at their respective resonant frequencies for the (a) conventional silicon waveguide, (c) symmetric slot-waveguide, and (e) asymmetric slot-waveguide. (b), (d), and (f) show the SER enhancement factors versus frequency for the corresponding finite-length ($\ell=250$~nm) waveguide structures of (a), (c), and (e), respectively. The vertical dashed lines in (b), (d), and (f) are at the resonant frequencies of the $\text{TM}_0$ (red) and the $\text{TM}_1$ (blue) mode calculated from the dispersion relation illustrated in Figs.~\ref{fig:Dispersion}(c, d).} \label{fig:ModeDistribution}
\end{figure}

To illustrate the physical mechanism underpinning PGAs and to remind the reader of the slot-waveguide concept, we begin by examining the dispersion relation of rectangular silicon waveguides, shown in Fig.~\ref{fig:Dispersion}(b). The dispersion equation for the eigenmodes of the slot-waveguide structures can be found in Ref.\citenum{almeida2004guiding} and \citenum{ma2009analysis}. The colored lines in Figs.~\ref{fig:Dispersion}(c, d) correspond to configurations without a gap, with a centered (along $z$) air gap and with an off-centered air gap, each 2 nm-thick. The width of the waveguides is set to 240 nm and the total height including the gap thickness is set to 582 nm. Propagation in the waveguide is along the $x$-direction with a propagation constant of $k_x$, i.e., along the direction normal to the cross-sections shown in the inset of Fig.~\ref{fig:Dispersion}(d). We consider transverse magnetic polarization, so that the electric field orientation ($E_z$) is perpendicular to the plane of the gap layer. Since the normal component of the electric displacement field, $\mathbf{D}=\epsilon\mathbf{E}$, is continuous at the two dielectric interfaces, the $E_z$ component of the quasi-$\text{TM}$ modes is stronger in the low-index region of the gap than in the high-index region of the waveguide. The field strength within the air gap layer is proportional to $E_{z,\text{air}}=(\epsilon_{\text{Si}}/\epsilon_{\text{air}})E_{z,\text{Si}}$ and can be further maximized by increasing the difference between the permittivity of the gap material and the waveguide material. In the spectral range of interest, we have two eigenmodes for the slot-waveguides, the $\text{TM}_0$ and $\text{TM}_1$ modes, originating from the even ($\text{TM}_0$) and odd ($\text{TM}_1$) modes of the unperturbed silicon waveguide. We note that the presence of the gap only weakly perturbs the dispersion relation of the silicon waveguide. In Figs.~\ref{fig:ModeDistribution}(a), (c), and (e), we plot the electric-field mode profiles in each of these waveguides at frequencies corresponding to the normalized propagation constant of $k_x \ell/2\pi=0.5$, where $\ell=250$~nm.

To realize PGAs, as in Fig.~\ref{fig:Dispersion}(a), the slot-waveguides must be truncated so as to satisfy the Fabry-Perot resonance condition $k_x \ell/2\pi=m/2$ for a positive integer value of $m$ and a finite length, $\ell$, along the propagation direction~\bibnote{This resonance condition is valid for waveguide modes with a near-zero reflection phase delay at the waveguide-air facets along the propagation direction. Otherwise, one should include both propagation phase delay and reflection phase delay.}. This condition is highlighted by a vertical dashed line in Figs.~\ref{fig:Dispersion}(c, d), which corresponds to the $m=1$, $\ell=250$~nm resonance condition. In Fig.~\ref{fig:ModeDistribution}(a), we show the mode profiles for the even ($\text{TM}_0$) and odd ($\text{TM}_1$) eigenmodes of the conventional silicon waveguide with no gap. For the symmetric slot-waveguide, shown in Fig.~\ref{fig:ModeDistribution}(c), the $E_z$ component of the $\text{TM}_0$ mode becomes strongly concentrated in the gap. From the dispersion relation in Fig.~\ref{fig:Dispersion}, we also see that the effective index ($n_{\text{eff}}$) for this mode slightly differs from that of the conventional silicon waveguide. In contrast, in the case of the $\text{TM}_1$ mode, the field profile and $n_{\text{eff}}$ values are indistinguishable compared to those of the conventional silicon waveguide. This is a consequence of the vanishing $E_{z,\text{Si}}$ component at the gap position. For the asymmetric slot-waveguide, both TM modes have different mode profiles and $n_{\text{eff}}$ values than the corresponding modes in the conventional silicon waveguide. From Fig.~\ref{fig:ModeDistribution}(e), we note that $E_{z,\text{Si}}$ for $\text{TM}_1$ mode of the asymmetric slot-waveguide has a non-zero value at the gap position and both $\text{TM}_0$ and $\text{TM}_1$ modes have their maximum field values located near the gap. Hence, in contrast to the symmetric case, an emitter embedded inside the gap of a resonating asymmetric slot-waveguide will strongly interact with both modes.

To evaluate the electric field confinement capability of the gap modes in the PGAs ($\ell=250$~nm), we calculate the quasinormal mode (QNM) volume, $V$, at the position of maximum field intensity of each localized mode. This occurs in the middle of the gap layer and 110 nm away from the $yz$-plane of symmetry along the $x$-axis. The quality factors (Q-factors) for both resonant modes of the PGAs and the conventional dielectric antenna are relatively low ($\text{Q}_{\text{TM}_0}\approx7$, $\text{Q}_{\text{TM}_1}\approx14$), which is to be expected due to their strong radiative leakage. The QNM formalism~\cite{sauvan2013theory, lalanne2018light} is used to calculate the mode volumes unambiguously, while also addressing the normalization issues~\cite{kristensen2012generalized} arising from the mode volume definition in leaky cavities. QNM theory yields complex mode volumes that are characteristic of non-hermitian resonators~\cite{cognee2019mapping}, however for the sake of simplicity, we neglect the imaginary part. The resonant $\text{TM}_0$ mode in the symmetric PGA has a mode volume of $\sim5\times 10^{-4} \lambda_{0}^3$, whereas the resonant $\text{TM}_0$ and $\text{TM}_1$ modes of the asymmetric PGA have mode volumes of $\sim13 \times 10^{-4} \lambda_{0}^3$ and $\sim9 \times 10^{-4} \lambda_{0}^3$, respectively, where $\lambda_{0}$ denotes the free-space wavelengths at corresponding resonant frequencies of the modes. In contrast, the $\text{TM}_0$ of the conventional dielectric antenna shows $V\approx 5\times10^{-2}\lambda_{0}^3$, which is 100 times smaller than the PGA. It is worth noting that the tight field confinement provided by PGAs is even on par with that provided by plasmonic antennas~\cite{biagioni2012nanoantennas, lalanne2019quasinormal}, but without ohmic losses.

\section{Results \& Discussion}

\subsection{Enhancement of Spontaneous Emission}

Resonant photonic structures can enhance an emitter's radiative decay rate by virtue of the spectral and spatial confinement of electromagnetic radiation. In the quantum picture, this change in SER comes about from a modification of the amplitude of vacuum field fluctuations at the position and orientation of the emitter, within its bandwidth. This is usually quantified by the change in the projected local density of optical states~\cite{barnes2020classical}. Classically, this effect can be understood as being due to the action of the scattered field due to the resonator, which acts back on the dipole (emitter) and can be quantified by measuring the change in input impedance of the dipole~\cite{barnes2020classical}. The SER enhancement factor is described by $\Gamma_{\text{r}}/\Gamma_0$, where $\Gamma_{\text{r}}$ is the radiative decay rate of the emitter in the presence of the resonating structure and $\Gamma_0$ is the decay rate of the same emitter in free-space. 

To study the SER enhancement of an emitter within the rectangular cross-section PGAs described in the previous section and to compare their enhancement capability with the conventional dielectric antenna without gap, we place a 2 nm long and infinitesimally thin current dipole at the position where the $E_{z}$ component is maximal for the resonating $\text{TM}_0$ mode of each structure. The dipole moment is oriented along the $z$-axis to maximize the interaction with the resonating modes. Finite-difference time-domain (FDTD) 3D simulations are used to calculate the total radiated power from the real part of the impedance, $\Re(Z)$, of the dipole~\cite{krasnok2015antenna}, i.e., $P=\frac{1}{2} |I|^2 \Re(Z)$, where $|I|$ is the amplitude of applied current in the dipole. In the absence of material loss, the non-radiative part of the decay rate is zero, and the SER enhancement factor becomes $\Gamma_{\text{r}}/\Gamma_0=P/P_{0}$, where $P$ and $P_{0}$ are the powers radiated by the dipole in the vicinity of the PGA and in free-space, respectively. In Figs.~\ref{fig:ModeDistribution}(b), (d), and (f), we plot the SER enhancement factors for the conventional dielectric antenna, the symmetric PGA and the asymmetric PGA, respectively. We find that the PGAs show a SER that is 2 orders of magnitude ($\times 10^2$) faster than for an emitter in the conventional Si dielectric antenna. This is a direct consequence of their reduced mode volumes. Compared to an emitter in free space, this corresponds to a SER enhancement by 3 orders of magnitude ($\times 10^3$). Furthermore, the frequencies corresponding to the peaks in SER enhancement agree well with the resonant frequencies calculated from the waveguide dispersion relation in Fig.~\ref{fig:Dispersion}. These are shown as vertical dashed lines in Figs.~\ref{fig:ModeDistribution}(b), (d), and (f).

\subsection{Introduction of Elliptical PGAs}

Although the rectangular pillars already highlight the fundamental features of PGAs, their performance can be further improved by modifying the cross-sectional shape. For example, by tapering the rectangular waveguide along the $y$-axis, the $n_{\text{eff}}$ for both modes can be further reduced to $\sim$1. This shifts the electromagnetic energy density from the high-index regions to the low-index regions and subsequently leads to increased spatial confinement in the gap. With this in mind, we replace the rectangular cross-sections with elliptical ones (with their major axis aligning the propagation direction of the initial rectangular slot-waveguides), as shown in Fig.~\ref{fig:StructurePurcellFbr}(a).

Elliptical PGAs with same height and gap position as the symmetric and asymmetric rectangular PGAs can increase the SER enhancement factor by more than $20\%$ for both resonant modes, as shown in Fig.~\ref{fig:StructurePurcellFbr}(b), while keeping the resonant frequencies the same. The resonant $\text{TM}_0$ mode in symmetric elliptical PGA and the resonant $\text{TM}_1$ mode in asymmetric elliptical PGA can provide a SER enhancement factor of $\sim$1300 and $\sim$1200, respectively. In addition, the gap position in asymmetric elliptical PGA allows the embedded dipole emitter to radiate efficiently over $\Delta\lambda$~\textgreater250 nm, via coupling to both resonant modes, with a SER enhancement factor \textgreater500. The SER enhancement factors for the variation in the lateral position of the emitter (along the $x$-axis, centered in the gap) is shown in Supplementary Fig.~\ref{fig:SER_lateral}. Firstly, we observe that even at the center of the PGA ($x=0$ \textmu m), where the $E_z$ component of the resonating $\text{TM}_0$ and $\text{TM}_1$ modes vanishes, there is still a considerable SER enhancement ($\Gamma_{\text{r}}/\Gamma_0\approx85$). This is due to the coupling of the emitter to the non-resonating part of the TM gap modes in the background~\cite{denning2018cavity}, and the SER enhancement factor is equivalent to the case of an emitter embedded in an infinite long slot-waveguide~\cite{jun2009broadband}. Secondly, the region of maximal enhancement in PGAs lies near the edges of the gap layer and therefore, a possible realization of the air-gap PGA with a glass support structure at the center (based on the fabrication techniques in Ref.~\citenum{lee2010silicon}, see Supplementary Fig.~\ref{fig:SER_fabrication}) does not impact the maximum SER achievable. In the following sections, we will restrict our attention to elliptical PGAs, due to their superior performance over rectangular PGAs.

\subsection{Radiation Pattern Engineering}

\begin{figure}[b]
\centering 
\includegraphics[width=0.85\textwidth]{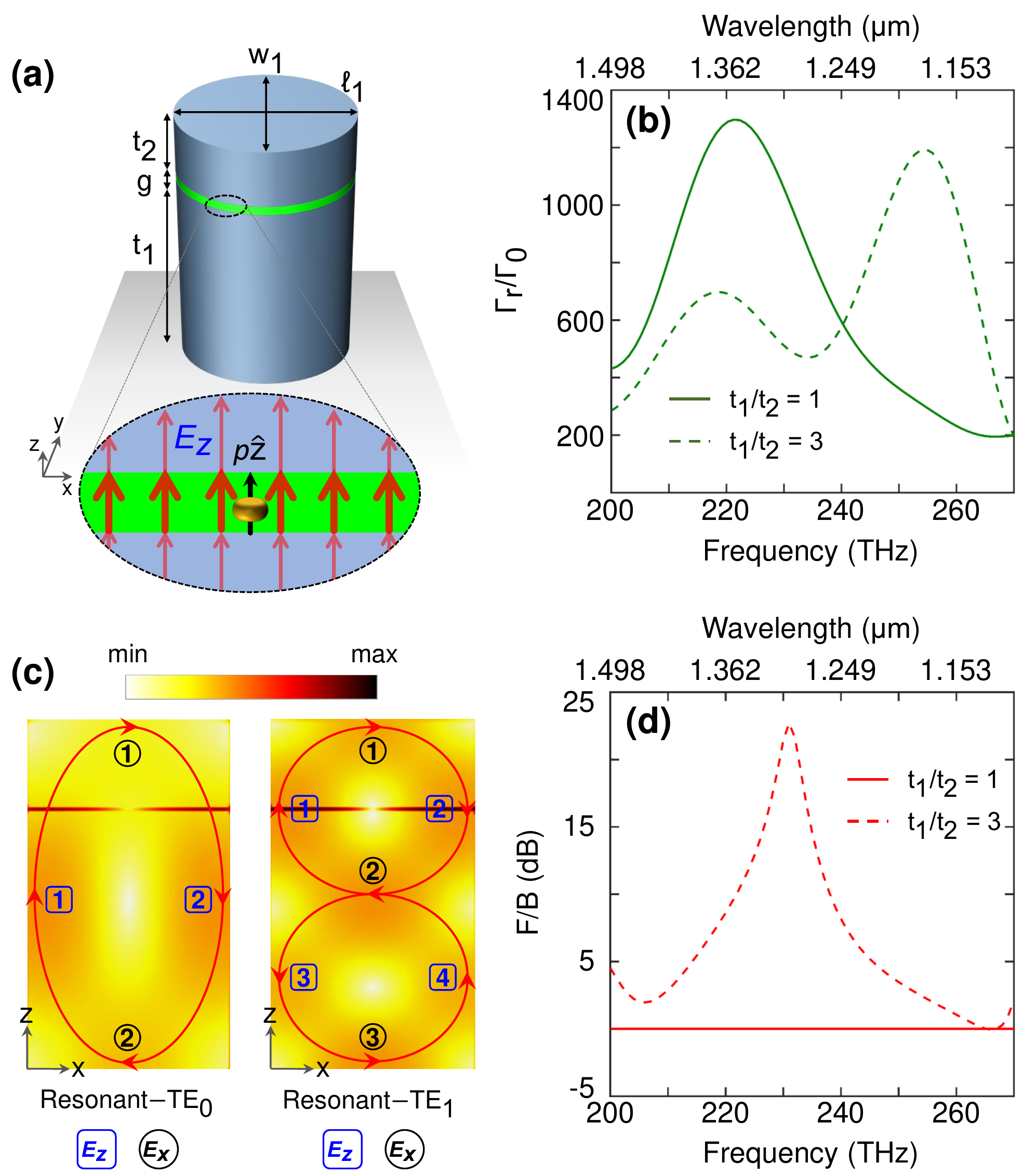}
\caption{Design and emission properties of elliptical PGAs. (a) Perspective view of the asymmetric elliptical PGA ($t_1/t_2=3$) with design parameters (in nm) $w_1=240$, $\ell_1=300$, $t_1=435$, $t_2=145$, and $g=2$~nm. The emitter is $\hat{\textbf{\emph{z}}}$ oriented and positioned within the gap layer 124 nm away from the vertical central axis of the structure along the $x$-axis. (b) SER enhancement factors versus frequencies for the asymmetric PGA in (a) and the symmetric PGA with $t_1/t_2=1$ ($t_1+t_2$ constant). (c) Electric field distribution in the central $xz$-plane of the asymmetric PGA for the $\text{TM}_0$ (left) and $\text{TM}_1$ (right) modes. (d) F/B ratio as a function of frequency for asymmetric and symmetric PGAs.} \label{fig:StructurePurcellFbr}
\end{figure}

The radiation pattern of PGAs primarily depends on the spatial and spectral overlap between the free-space radiation modes and the waveguide modes on each facet of the structure. The resonant $\text{TM}_0$ and $\text{TM}_1$ modes are zero-order modes along the $y$-axis, and the corresponding mode profiles have a single antinode in this direction. Therefore, we can restrict our attention to a 2D slice in the $xz$-central plane of the antenna to understand the radiation pattern. In this plane, both $E_z$ and $E_x$ components of the resonating $\text{TM}_0$ mode have a single antinode (even parity), as shown in Fig.~\ref{fig:StructurePurcellFbr}(c) in the case of asymmetric PGAs ($t_1/t_2=3$). Such a field distribution suggests that the resonating $\text{TM}_0$ mode can couple to plane waves propagating along the $x$- and $z$-axes ($E_z$:~\squared{1} and \squared{2} ---responsible for radiation in the $-x$-direction and the $+x$-direction, respectively. $E_x$:~\circled{1} and \circled{2} ---responsible for radiation in the $+z$-direction and the $-z$-direction, respectively.). In contrast, the $E_x$ components of the resonating $\text{TM}_1$ mode have a single antinode and the $E_z$ components have two antinodes of opposing sign (odd parity). Since these antinodes of the $E_z$ components are within the silicon, they are spaced by a distance smaller than the half-wavelength in air and result in destructive interference along the $x$-axis ($E_z$:~\squared{1} and \squared{3} ---responsible for null-radiation in the $-x$-direction. $E_z$:~\squared{2} and \squared{4} ---responsible for null-radiation in the $+x$-direction). This maximizes the radiation of resonant $\text{TM}_1$ mode along the $z$-axis due to coupling of the $E_x$ components of the mode with propagating plane waves in that direction ($E_x$:~\circled{1} ---responsible for radiation in the $+z$-direction, $E_x$:~\circled{3} ---responsible for radiation in the $-z$-direction). 

To maximize the collection efficiency of a nearby free-space optical system, PGAs should ideally radiate unidirectionally out-of-plane (here in the $-z$-direction). The symmetric PGAs behave as single mode resonators supporting the $\text{TM}_0$ mode with negligible coupling of the emitter to the $\text{TM}_1$ mode and leak a significant amount of radiation along the $\pm x$-directions in addition to the desired $\pm z$-directions, therefore, they produce an omnidirectional radiation pattern. Offsetting the position of the gap allows for directionality. First, this allows coupling of the emitter to the $\text{TM}_1$ mode, which ensures predominant radiation along the $\pm z$-directions while minimizing the radiation along the $\pm x$-directions at the resonant frequency of the $\text{TM}_1$ mode. Second, the asymmetric gap position leads to unequal perturbations of the $E_x$ components of the modes along the $z$-axis and realizes higher directionality in a preferred direction. For example, the asymmetric PGA, shown in Figs.~\ref{fig:StructurePurcellFbr}(a, c), allows for strong excitation of the resonant $\text{TM}_1$ mode. Additionally, the modal field distributions show a relatively strong $E_x$ component near the bottom air-dielectric interface ($E_x$:~\circled{2} and \circled{3} in the case of resonant $\text{TM}_0$ and $\text{TM}_1$, respectively.) compared to the one near the top air-dielectric interface ($E_x$:~\circled{1} in the case of both resonant $\text{TM}_0$ and $\text{TM}_1$). To quantify the asymmetric radiation of PGAs along the $z$-axis, we calculate the ratio of the power radiated along the $-z$-direction (forward; F) to the power radiated along the $+z$-direction (backward; B)---the F/B ratio. For the asymmetric PGA ($t_{1}/t_{2}=3$), we observe a F/B ratio \textgreater$3$~dB, as shown in Fig.~\ref{fig:StructurePurcellFbr}(d), for both the $\text{TM}_0$ and $\text{TM}_1$ mode at their respective resonant frequencies~\bibnote{It should be noted that the emitter is at the position of peak field intensity for both resonant modes of the structure. Hence, the dipole mostly decays through coupling to the antenna modes and negligibly through direct coupling to free-space modes.}.

\begin{figure}[b]
\centering 
\includegraphics[width=0.65\textwidth]{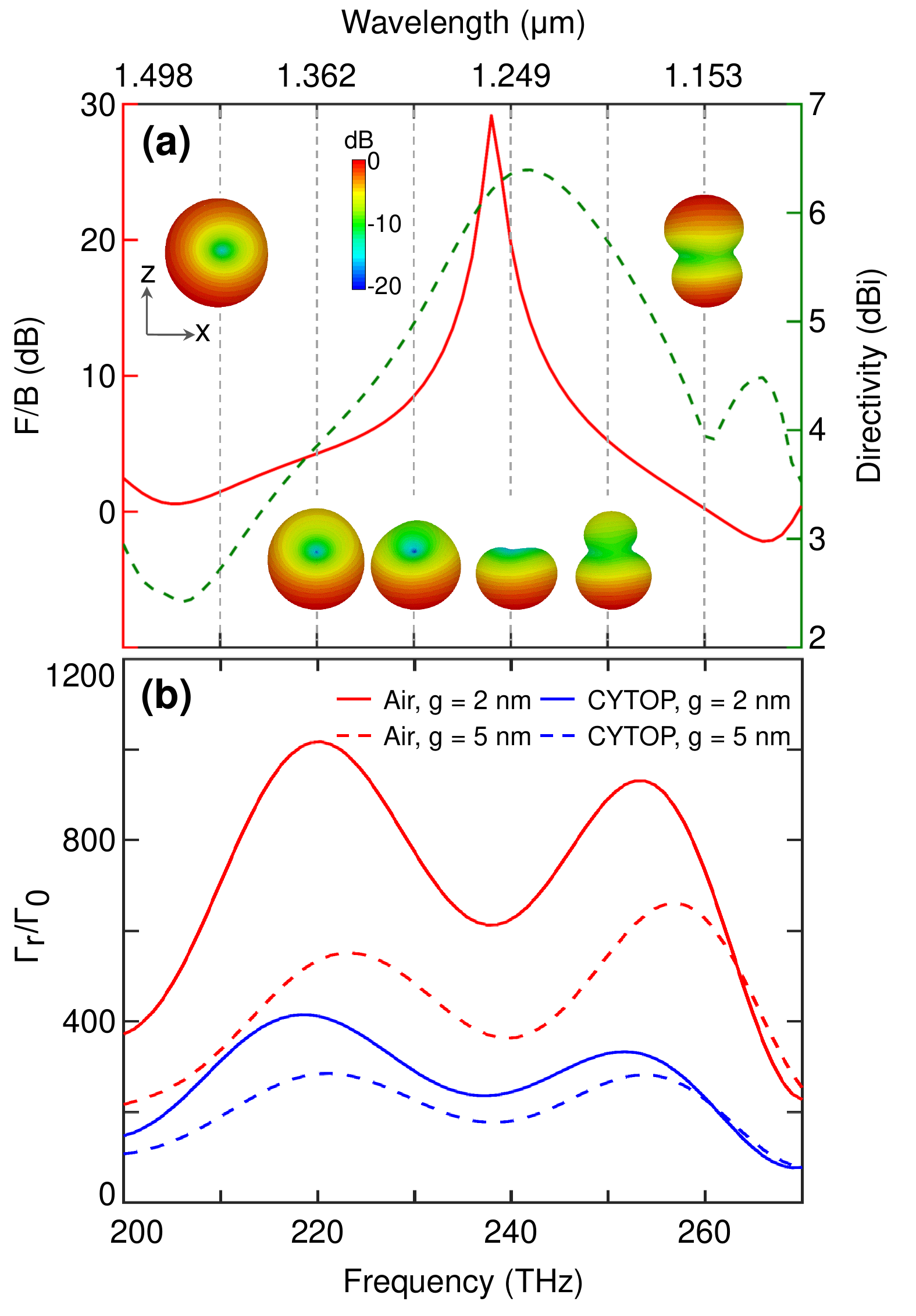}
\caption{Radiation properties of an asymmetric PGA with $t1/t2=1.76$ and SER enhancement factors for different gap thicknesses and materials. The total thickness of silicon ($t_1+t_2$), $w_1$ and $\ell_1$ are same as in Fig.~\ref{fig:StructurePurcellFbr}(a). (a) F/B ratio (left axis) and directivity (right axis) versus frequency for the PGA with a 2 nm air gap. The insets show the radiation patterns of the PGA in the $xz$-plane at the corresponding frequencies with their maximum directivity normalized to 0 dB. (b) SER enhancement factors versus frequency for the PGA with air gaps (red) and CYTOP gaps (blue) of thickness $g=2$ nm (solid lines) and $g=5$ nm (dashed lines).} \label{fig:improvedbandwidthdirectionality}
\end{figure}

Because of the low-$Q$ characterizing both resonant modes, emitters in the asymmetric PGAs can readily excite both simultaneously. For an emitter frequency in between the resonant frequencies of both antenna modes, we observe a F/B ratio of 22 dB, as shown in Fig.~\ref{fig:StructurePurcellFbr}(d), due to constructive interference along the $-z$-direction and destructive interference along the $+z$-direction. This phenomenon of directional radiation due to interference between modes has been studied extensively for Mie resonators such as silicon nanospheres, where the Kerker condition between electric and magnetic dipoles of the structure leads to unidirectional emission~\cite{warne2012perturbation, staude2013tailoring, liu2018generalized}. It turns out that the asymmetric PGA design with $t_{1}/t_{2}=3$ offer a relatively low SER enhancement factor in the frequency range of maximum F/B ratio. Fortunately, this can be improved by tuning the coupling-strength of the emitter with the resonating $\text{TM}_0$ and $\text{TM}_1$ mode through the choice of the gap position. For example, Fig.~\ref{fig:improvedbandwidthdirectionality} shows results for an asymmetric PGA with $t_{1}/t_{2}=1.76$ that can attain a SER enhancement factor of $\sim$600 at an intermediate frequency, while radiating with a F/B ratio of 30 dB. This strategy also improves the spectral bandwidth of the PGA to $\Delta\lambda=315$~nm at $\lambda=1.25$ \textmu m. Note that the resonant frequencies of the structure remain unchanged when moving the gap to this new position due to the small changes in the corresponding $n_{\text{eff}}$ of the slot-waveguide modes. This is in contrast to strategies where dielectric structures must be optimized to place both resonant frequencies close to each other to achieve directionality~\cite{warne2012perturbation, staude2013tailoring}.

The superposition of the $\text{TM}_0$ and $\text{TM}_1$ modes also allows for radiation leakage in the $\pm x$-direction. For this purpose, it is essential to consider the antenna directivity in addition to its F/B ratio. In Fig.~\ref{fig:improvedbandwidthdirectionality}(a), we plot the directivity of the PGA (with $t_{1}/t_{2}=1.76$) normalized to that of an isotropic antenna~\cite{balanis2016antenna}. We obtain a maximum directivity of $\sim$6.2 dBi near the frequency range of the maximum F/B ratio. The radiation pattern of this PGA in the $xz$-plane (normalized to 0 dB) is shown at selected frequencies in the inset of Fig.~\ref{fig:improvedbandwidthdirectionality}(a). These results demonstrate that PGA can radiate directionally over a broad bandwidth towards the collection optics in the bottom ($-z$-direction). To change the maximum radiation direction towards the top ($+z$-direction), the antenna structure simply needs to be flipped upside-down. 

\subsection{Influence of Gap Thickness \& Material}

The performance of PGAs is primarily governed by the modal distribution near the gap region. This distribution can be tuned by varying the gap thickness, and the index-contrast between the gap material and the pillar material. To study the impact of the parameters, we show additional results for PGA designs with a gap thickness of 5 nm or the use of the polymer CYTOP ($n=1.33$) as the gap material. To first order, the field in the sub-wavelength dielectric gap is homogeneous and when the gap size increase, the electromagnetic energy shifts to the higher index region, which increases the mode volumes. As shown in Fig.~\ref{fig:improvedbandwidthdirectionality}(b), thicker gaps and lower index-contrast both decrease the SER enhancement. We observe a SER enhancement that is $\sim$2.5 times smaller ($\Gamma_{\text{air}}/\Gamma_{\text{CYTOP}}$) for a 2 nm CYTOP gap as compared to that of an air gap. In contrast, the $\Gamma_{\text{air}}/\Gamma_{\text{CYTOP}}$ becomes $\sim$2 in the case of a 5 nm gap. This shows that the impact of gap thickness on SER is reduced for lower index contrasts. We also study the influence of a glass substrate ($n=1.50$) on the SER enhancement factor of PGAs with different gap thickness and materials, and present the results in Supplementary Fig.~\ref{fig:SER_substrate}. We note that due to substrate-induced asymmetry, both the SER enhancement factor and the F/B ratio can be improved further in the frequency range between the resonances.

\subsection{PGAs as Receivers}

Light reception is the reciprocal of emission. As receivers, dielectric optical antennas collect radiation, which is then converted to oscillating electric and magnetic fields confined in the near-field. The collection efficiency of an antenna can be characterized by its extinction cross-section, whereas the conversion and concentration efficiencies can be characterized by the localized field (amplitude) enhancement or intensity enhancement factor. To study these figures of merit for PGAs, we consider a linearly $x$-polarized plane wave incident on the PGAs. Since the PGA is designed to efficiently radiate in the $-z$-direction, the propagation of the incident plane wave is chosen in the reciprocal direction ($+z$). Here, we report results for the symmetric and asymmetric PGAs ($t_{1}/t_{2}:1,  1.76, 3$) with 2 and 5~nm thick CYTOP gaps ($g=2, 5$). The corresponding results for the PGAs with air gaps are presented in Supplementary material (Fig.~\ref{fig:CrosssecFieldenahnceWithair}).

The collection efficiency of an antenna relates the amount of power received, $P_{\text{r}}$, by the antenna to the incident plane wave power density, $S$, i.e., $P_{\text{r}}/S$. The received power can be expressed by $P_{\text{r}}=S A_{\text{eff}} T$, where $A_{\text{eff}}$ is the effective aperture of the antenna in the direction of the incoming plane wave and $T$ is the intensity transmission coefficient~\cite{pursula2007antenna}. The parameter, $A_{\text{eff}}=\lambda^{2}D/4\pi$, is a  measure of normalized directivity, where $D$ is the directivity of the antenna when receiving in the direction of plane wave propagation. In the case of nanoscale antennas, measuring the received power is difficult from the antenna end, however, the scattered power, $P_{\text{sca}}$, can be measured instead. In the absence of ohmic losses, the scattered power is the same as the received power  ($P_{\text{r}}=P_{\text{sca}}+P_{\text{abs}}; P_{\text{abs}}=0$). Therefore, the collection efficiency, $P_{\text{r}}/S$, of the antenna can be characterized by the extinction cross-section (here the same as the scattering cross-section), $\sigma_{\text{ext}}=\sigma_{\text{sca}}=P_{\text{sca}}/S$. At the resonant frequencies of the structure, $T \to 1$, and $\sigma_{\text{ext}}$ essentially depend on the $A_{\text{eff}} \propto D$ of the antenna. 

\begin{figure}[b]
\centering 
\includegraphics[width=0.85\textwidth]{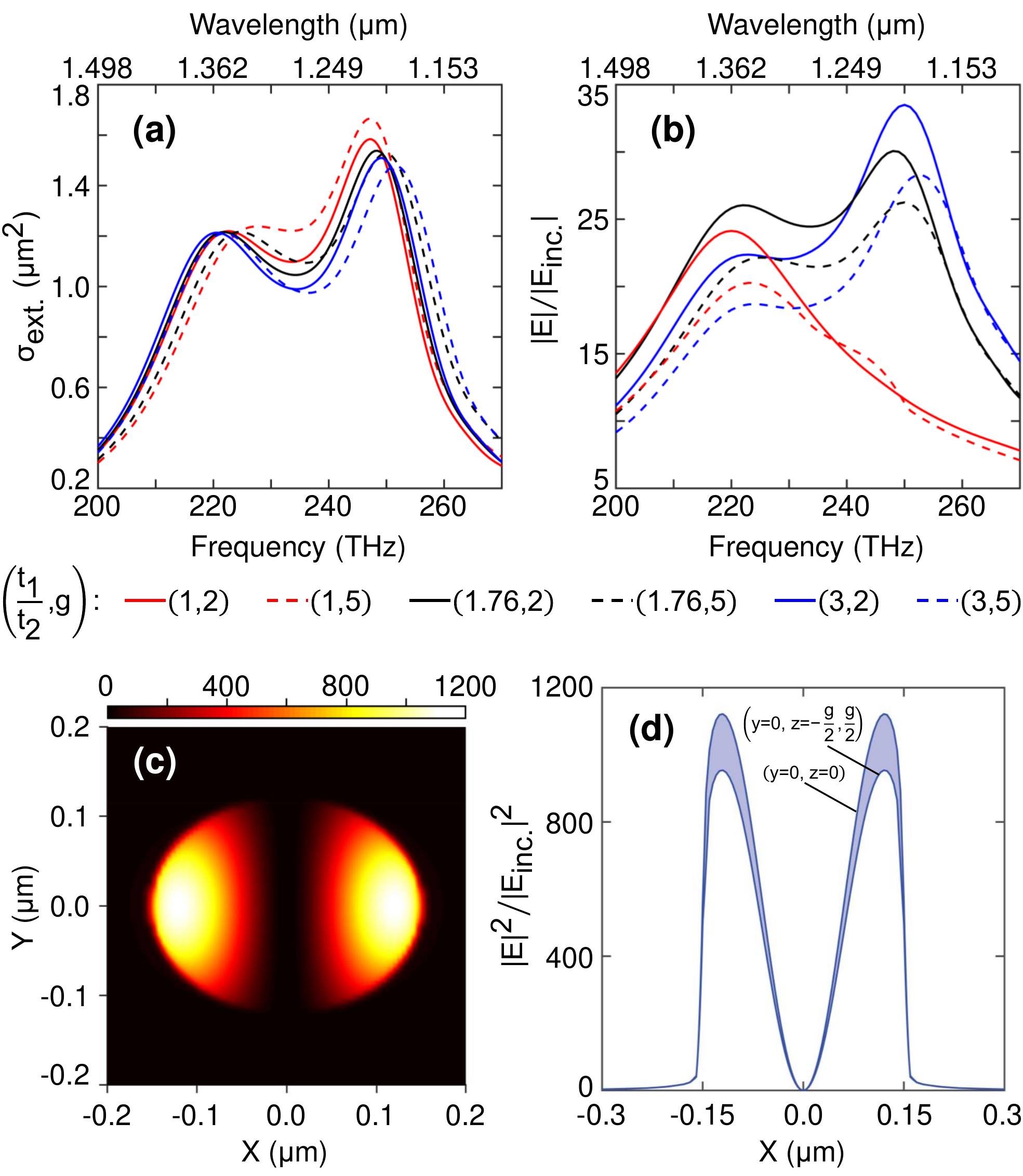}
\caption{Extinction cross-sections and electric field enhancement factor of PGAs with CYTOP gap. (a) Extinction cross-section ($\sigma_{\text{ext.}}=\sigma_{\text{sca.}}$) versus frequency, and (b) field enhancement factor $|\bf{E}|/|\bf{E}_\text{inc.}|$ as a function of  frequency for elliptical PGAs, where the gap position and thickness are varied. The combined thickness of silicon ($t_1+t_2$), $w_1$ and $\ell_1$ are the same as for the PGA in Fig.~\ref{fig:StructurePurcellFbr}(a). (c) Near-field intensity distribution for the asymmetric PGA ($t_1/t_2=3$) with $g=2$~nm in a plane parallel to $xy$-plane and passing through the center of the gap layer, and (d) intensity distribution within the gap layer along $x$-axis at $y=0$. The upper line of the shaded region represents the intensity profile at the center of the gap ($z=0$), whereas the lower bound represents the intensity profile near the CYTOP-Si boundaries ($z=-g/2,g/2$). The field intensity increases from the bottom CYTOP-Si boundary to the center of the gap and decreases in a similar pattern up to the top CYTOP-Si boundary.} \label{fig:CrosssecFieldenahnceWithcytop}
\end{figure}

As shown in Fig.~\ref{fig:CrosssecFieldenahnceWithcytop}(a), we observe a relatively high $\sigma_\text{ext}$ at the resonant frequency of the $\text{TM}_1$ mode as compared to that of the $\text{TM}_0$ mode. This is because the $\text{TM}_1$ mode of the PGA has a higher directionality (bidirectional) in the $-z$-direction compared to the $\text{TM}_0$ mode (omnidirectional)~\bibnote{The spectral directivity values of PGAs in the receiving configuration are similar to those of transmission. However, both values differ from each other when the emitter is weakly coupled to the resonant modes of the structure. For an accurate characterization of directivity during reception, the resonant modes should be excited by a plane wave propagating towards the structure in the desired direction.}. At frequencies where the antenna is off-resonance, the amount of received power ($P_{\text{r}}$) decreases due to the lower transmission coefficient ($T<<1$). Hence, in the frequency range between the two resonances of the structure, $\sigma_{\text{ext}}$ decreases, in spite of the high directionality ($D$). A parameter closely related to $\sigma_{\text{sca}}$ ($=\sigma_{\text{ext}}$) is the scattering efficiency $Q_\text{sca}=\sigma_{\text{sca}}/C_{\text{g}}$, which can be calculated for a geometric cross-section $C_{\text{g}}=0.0565$ \textmu$\text{m}^2$ of the PGA. We obtain a maximum $Q_{\text{sca}}$ of $\sim$30 for the case of symmetric PGA ($t_{1}/t_{2}=1$) with a 5 nm CYTOP gap.

The near-field enhancement factor is defined as the ratio between the maximum electric field amplitude in the vicinity of the PGA to the amplitude of the incident plane wave. Higher values of the field enhancement factor are obtained when the PGA receives more power ($\propto S A_{\text{eff}} T$) and stores it in a smaller three-dimensional space for a longer time. In particular, symmetric PGAs ($t_{1}/t_{2}=1$) receive a relatively high amount of power at the resonant frequency of the $\text{TM}_1$ mode. However, this mode is a bulk mode (the same as that of the conventional dielectric pillar without a gap) and fails to store the energy within the gap. As shown in Fig.~\ref{fig:CrosssecFieldenahnceWithcytop}(b), this phenomenon leads to relatively low field enhancements at the respective resonant frequencies. In contrast, the very same mode in asymmetric PGAs ($t_{1}/t_{2}=3$) can strongly confine the near-field energy in the gap region and can have field enhancement factors as high as $\sim$35 (intensity enhancement factor: $\sim$1200) for a 2 nm CYTOP gap and $\sim$55 (intensity enhancement factor: $\sim$3000) for an air gap of the same thickness (see Supplementary Fig.~\ref{fig:CrosssecFieldenahnceWithair} for PGAs with air gaps). As expected, this value decreases with increasing value of the gap thickness. Similar to the transmission configuration, the receiving configuration of asymmetric PGAs allows for high field enhancement over a broad spectral bandwidth of $\Delta \lambda \approx300$ nm. For the asymmetric PGA ($t_{1}/t_{2}=3, g= 2$~nm), we show in Fig.~\ref{fig:CrosssecFieldenahnceWithcytop}(c) the near-field intensity distribution in a plane parallel to the $xy$-plane and passing through the center of the gap. We observe an essentially uniform distribution of intensity along the gap thickness from the intensity distribution plot shown in Fig.~\ref{fig:CrosssecFieldenahnceWithcytop}(d).

\section{Conclusion}

We have presented a new all-dielectric optical antenna that can serve as a platform for engineering light-matter interaction on par with that of plasmonic antennas. PGAs exploit the properties of deeply subwavelength slot-waveguide modes both for strong spatial confinement and for unidirectional radiation over a broad spectrum. We have shown that the radiated power of a quantum emitter in free-space can be improved by \textgreater$1000$ times by embedding it within the gap of PGA, while maintaining a quantum efficiency of $\sim$100$\%$. The use of all-dielectric structures rather than metallic ones becomes of increasing importance for applications that are sensitive to heating or require high quantum efficiency. In the receiving regime, PGAs can tightly confine the field of an incident plane wave with an intensity enhancement of up to $\sim$3000. This can be useful both for interacting with localized emitters and for enhancing non-linear effects in cases where heating from plasmonic antennas can be problematic. These results are particularly compelling given the relative simplicity of the antenna structure. Our study could be further extended to improve these enhancement factors as well as the radiation directionality of PGAs by fabricating the antennas on a dielectric mirror or numerically optimizing the PGA cross-section.

\section{Acknowledgements}

This work was supported by the Natural Sciences and Engineering Council of Canada Strategic Grant program and the Canada Research Chairs program. The work at CUNY was supported by NSF QII-TAQS grant
\#1936351.

\bibliography{achemso-demo}

\providecommand{\latin}[1]{#1}
\makeatletter
\providecommand{\doi}
  {\begingroup\let\do\@makeother\dospecials
  \catcode`\{=1 \catcode`\}=2 \doi@aux}
\providecommand{\doi@aux}[1]{\endgroup\texttt{#1}}
\makeatother
\providecommand*\mcitethebibliography{\thebibliography}
\csname @ifundefined\endcsname{endmcitethebibliography}
  {\let\endmcitethebibliography\endthebibliography}{}
\begin{mcitethebibliography}{54}
\providecommand*\natexlab[1]{#1}
\providecommand*\mciteSetBstSublistMode[1]{}
\providecommand*\mciteSetBstMaxWidthForm[2]{}
\providecommand*\mciteBstWouldAddEndPuncttrue
  {\def\EndOfBibitem{\unskip.}}
\providecommand*\mciteBstWouldAddEndPunctfalse
  {\let\EndOfBibitem\relax}
\providecommand*\mciteSetBstMidEndSepPunct[3]{}
\providecommand*\mciteSetBstSublistLabelBeginEnd[3]{}
\providecommand*\EndOfBibitem{}
\mciteSetBstSublistMode{f}
\mciteSetBstMaxWidthForm{subitem}{(\alph{mcitesubitemcount})}
\mciteSetBstSublistLabelBeginEnd
  {\mcitemaxwidthsubitemform\space}
  {\relax}
  {\relax}

\bibitem[Novotny and Van~Hulst(2011)Novotny, and
  Van~Hulst]{novotny2011antennas}
Novotny,~L.; Van~Hulst,~N. Antennas for light. \emph{Nat. Photonics}
  \textbf{2011}, \emph{5}, 83--90\relax
\mciteBstWouldAddEndPuncttrue
\mciteSetBstMidEndSepPunct{\mcitedefaultmidpunct}
{\mcitedefaultendpunct}{\mcitedefaultseppunct}\relax
\EndOfBibitem
\bibitem[Khurgin and Boltasseva(2012)Khurgin, and
  Boltasseva]{khurgin2012reflecting}
Khurgin,~J.~B.; Boltasseva,~A. Reflecting upon the losses in plasmonics and
  metamaterials. \emph{MRS Bull.} \textbf{2012}, \emph{37}, 768--779\relax
\mciteBstWouldAddEndPuncttrue
\mciteSetBstMidEndSepPunct{\mcitedefaultmidpunct}
{\mcitedefaultendpunct}{\mcitedefaultseppunct}\relax
\EndOfBibitem
\bibitem[Tsakmakidis \latin{et~al.}(2016)Tsakmakidis, Boyd, Yablonovitch, and
  Zhang]{tsakmakidis2016large}
Tsakmakidis,~K.~L.; Boyd,~R.~W.; Yablonovitch,~E.; Zhang,~X. Large
  spontaneous-emission enhancements in metallic nanostructures: towards LEDs
  faster than lasers. \emph{Opt. Express} \textbf{2016}, \emph{24},
  17916--17927\relax
\mciteBstWouldAddEndPuncttrue
\mciteSetBstMidEndSepPunct{\mcitedefaultmidpunct}
{\mcitedefaultendpunct}{\mcitedefaultseppunct}\relax
\EndOfBibitem
\bibitem[Koenderink(2017)]{koenderink2017single}
Koenderink,~A.~F. Single-photon nanoantennas. \emph{ACS Photonics}
  \textbf{2017}, \emph{4}, 710--722\relax
\mciteBstWouldAddEndPuncttrue
\mciteSetBstMidEndSepPunct{\mcitedefaultmidpunct}
{\mcitedefaultendpunct}{\mcitedefaultseppunct}\relax
\EndOfBibitem
\bibitem[Grober \latin{et~al.}(1997)Grober, Schoelkopf, and
  Prober]{grober1997optical}
Grober,~R.~D.; Schoelkopf,~R.~J.; Prober,~D.~E. Optical antenna: Towards a
  unity efficiency near-field optical probe. \emph{Appl. Phys. Lett.}
  \textbf{1997}, \emph{70}, 1354--1356\relax
\mciteBstWouldAddEndPuncttrue
\mciteSetBstMidEndSepPunct{\mcitedefaultmidpunct}
{\mcitedefaultendpunct}{\mcitedefaultseppunct}\relax
\EndOfBibitem
\bibitem[Bharadwaj \latin{et~al.}(2009)Bharadwaj, Deutsch, and
  Novotny]{bharadwaj2009optical}
Bharadwaj,~P.; Deutsch,~B.; Novotny,~L. Optical antennas. \emph{Adv. Opt.
  Photonics} \textbf{2009}, \emph{1}, 438--483\relax
\mciteBstWouldAddEndPuncttrue
\mciteSetBstMidEndSepPunct{\mcitedefaultmidpunct}
{\mcitedefaultendpunct}{\mcitedefaultseppunct}\relax
\EndOfBibitem
\bibitem[Kauranen and Zayats(2012)Kauranen, and Zayats]{kauranen2012nonlinear}
Kauranen,~M.; Zayats,~A.~V. Nonlinear plasmonics. \emph{Nat. Photonics}
  \textbf{2012}, \emph{6}, 737--748\relax
\mciteBstWouldAddEndPuncttrue
\mciteSetBstMidEndSepPunct{\mcitedefaultmidpunct}
{\mcitedefaultendpunct}{\mcitedefaultseppunct}\relax
\EndOfBibitem
\bibitem[Biagioni \latin{et~al.}(2012)Biagioni, Huang, and
  Hecht]{biagioni2012nanoantennas}
Biagioni,~P.; Huang,~J.-S.; Hecht,~B. Nanoantennas for visible and infrared
  radiation. \emph{Rep. Prog. Phys.} \textbf{2012}, \emph{75}, 024402\relax
\mciteBstWouldAddEndPuncttrue
\mciteSetBstMidEndSepPunct{\mcitedefaultmidpunct}
{\mcitedefaultendpunct}{\mcitedefaultseppunct}\relax
\EndOfBibitem
\bibitem[Agio and Al{\`u}(2013)Agio, and Al{\`u}]{agio2013optical}
Agio,~M.; Al{\`u},~A. \emph{Optical antennas}; Cambridge University Press,
  2013\relax
\mciteBstWouldAddEndPuncttrue
\mciteSetBstMidEndSepPunct{\mcitedefaultmidpunct}
{\mcitedefaultendpunct}{\mcitedefaultseppunct}\relax
\EndOfBibitem
\bibitem[Kuhlicke \latin{et~al.}(2013)Kuhlicke, Schietinger, Matyssek, Busch,
  and Benson]{kuhlicke2013situ}
Kuhlicke,~A.; Schietinger,~S.; Matyssek,~C.; Busch,~K.; Benson,~O. In situ
  observation of plasmon tuning in a single gold nanoparticle during controlled
  melting. \emph{Nano Lett.} \textbf{2013}, \emph{13}, 2041--2046\relax
\mciteBstWouldAddEndPuncttrue
\mciteSetBstMidEndSepPunct{\mcitedefaultmidpunct}
{\mcitedefaultendpunct}{\mcitedefaultseppunct}\relax
\EndOfBibitem
\bibitem[Caldarola \latin{et~al.}(2015)Caldarola, Albella, Cort{\'e}s, Rahmani,
  Roschuk, Grinblat, Oulton, Bragas, and Maier]{caldarola2015non}
Caldarola,~M.; Albella,~P.; Cort{\'e}s,~E.; Rahmani,~M.; Roschuk,~T.;
  Grinblat,~G.; Oulton,~R.~F.; Bragas,~A.~V.; Maier,~S.~A. Non-plasmonic
  nanoantennas for surface enhanced spectroscopies with ultra-low heat
  conversion. \emph{Nat. Commun.} \textbf{2015}, \emph{6}, 1--8\relax
\mciteBstWouldAddEndPuncttrue
\mciteSetBstMidEndSepPunct{\mcitedefaultmidpunct}
{\mcitedefaultendpunct}{\mcitedefaultseppunct}\relax
\EndOfBibitem
\bibitem[Mahmoudi \latin{et~al.}(2014)Mahmoudi, Lohse, Murphy, Fathizadeh,
  Montazeri, and Suslick]{mahmoudi2014variation}
Mahmoudi,~M.; Lohse,~S.~E.; Murphy,~C.~J.; Fathizadeh,~A.; Montazeri,~A.;
  Suslick,~K.~S. Variation of protein corona composition of gold nanoparticles
  following plasmonic heating. \emph{Nano Lett.} \textbf{2014}, \emph{14},
  6--12\relax
\mciteBstWouldAddEndPuncttrue
\mciteSetBstMidEndSepPunct{\mcitedefaultmidpunct}
{\mcitedefaultendpunct}{\mcitedefaultseppunct}\relax
\EndOfBibitem
\bibitem[Alessandri and Lombardi(2016)Alessandri, and
  Lombardi]{alessandri2016enhanced}
Alessandri,~I.; Lombardi,~J.~R. Enhanced Raman scattering with dielectrics.
  \emph{Chem. Rev.} \textbf{2016}, \emph{116}, 14921--14981\relax
\mciteBstWouldAddEndPuncttrue
\mciteSetBstMidEndSepPunct{\mcitedefaultmidpunct}
{\mcitedefaultendpunct}{\mcitedefaultseppunct}\relax
\EndOfBibitem
\bibitem[Barnes \latin{et~al.}(2002)Barnes, Bj{\"o}rk, G{\'e}rard, Jonsson,
  Wasey, Worthing, and Zwiller]{barnes2002solid}
Barnes,~W.; Bj{\"o}rk,~G.; G{\'e}rard,~J.; Jonsson,~P.; Wasey,~J.;
  Worthing,~P.; Zwiller,~V. Solid-state single photon sources: light collection
  strategies. \emph{Eur. Phys. J. D} \textbf{2002}, \emph{18}, 197--210\relax
\mciteBstWouldAddEndPuncttrue
\mciteSetBstMidEndSepPunct{\mcitedefaultmidpunct}
{\mcitedefaultendpunct}{\mcitedefaultseppunct}\relax
\EndOfBibitem
\bibitem[Kuznetsov \latin{et~al.}(2016)Kuznetsov, Miroshnichenko, Brongersma,
  Kivshar, and Luk’yanchuk]{kuznetsov2016optically}
Kuznetsov,~A.~I.; Miroshnichenko,~A.~E.; Brongersma,~M.~L.; Kivshar,~Y.~S.;
  Luk’yanchuk,~B. Optically resonant dielectric nanostructures.
  \emph{Science} \textbf{2016}, \emph{354}\relax
\mciteBstWouldAddEndPuncttrue
\mciteSetBstMidEndSepPunct{\mcitedefaultmidpunct}
{\mcitedefaultendpunct}{\mcitedefaultseppunct}\relax
\EndOfBibitem
\bibitem[Bouchet \latin{et~al.}(2016)Bouchet, Mivelle, Proust, Gallas, Ozerov,
  Garcia-Parajo, Gulinatti, Rech, De~Wilde, Bonod, \latin{et~al.}
  others]{bouchet2016enhancement}
Bouchet,~D.; Mivelle,~M.; Proust,~J.; Gallas,~B.; Ozerov,~I.;
  Garcia-Parajo,~M.~F.; Gulinatti,~A.; Rech,~I.; De~Wilde,~Y.; Bonod,~N.,
  \latin{et~al.}  Enhancement and inhibition of spontaneous photon emission by
  resonant silicon nanoantennas. \emph{Phys. Rev. Appl.} \textbf{2016},
  \emph{6}, 064016\relax
\mciteBstWouldAddEndPuncttrue
\mciteSetBstMidEndSepPunct{\mcitedefaultmidpunct}
{\mcitedefaultendpunct}{\mcitedefaultseppunct}\relax
\EndOfBibitem
\bibitem[Bozhevolnyi and Khurgin(2016)Bozhevolnyi, and
  Khurgin]{bozhevolnyi2016fundamental}
Bozhevolnyi,~S.~I.; Khurgin,~J.~B. Fundamental limitations in spontaneous
  emission rate of single-photon sources. \emph{Optica} \textbf{2016},
  \emph{3}, 1418--1421\relax
\mciteBstWouldAddEndPuncttrue
\mciteSetBstMidEndSepPunct{\mcitedefaultmidpunct}
{\mcitedefaultendpunct}{\mcitedefaultseppunct}\relax
\EndOfBibitem
\bibitem[Rutckaia \latin{et~al.}(2017)Rutckaia, Heyroth, Novikov, Shaleev,
  Petrov, and Schilling]{rutckaia2017quantum}
Rutckaia,~V.; Heyroth,~F.; Novikov,~A.; Shaleev,~M.; Petrov,~M.; Schilling,~J.
  Quantum dot emission driven by Mie resonances in silicon nanostructures.
  \emph{Nano Lett.} \textbf{2017}, \emph{17}, 6886--6892\relax
\mciteBstWouldAddEndPuncttrue
\mciteSetBstMidEndSepPunct{\mcitedefaultmidpunct}
{\mcitedefaultendpunct}{\mcitedefaultseppunct}\relax
\EndOfBibitem
\bibitem[Albella \latin{et~al.}(2013)Albella, Poyli, Schmidt, Maier, Moreno,
  S{\'a}enz, and Aizpurua]{albella2013low}
Albella,~P.; Poyli,~M.~A.; Schmidt,~M.~K.; Maier,~S.~A.; Moreno,~F.;
  S{\'a}enz,~J.~J.; Aizpurua,~J. Low-loss electric and magnetic field-enhanced
  spectroscopy with subwavelength silicon dimers. \emph{J. Phys. Chem. C}
  \textbf{2013}, \emph{117}, 13573--13584\relax
\mciteBstWouldAddEndPuncttrue
\mciteSetBstMidEndSepPunct{\mcitedefaultmidpunct}
{\mcitedefaultendpunct}{\mcitedefaultseppunct}\relax
\EndOfBibitem
\bibitem[Regmi \latin{et~al.}(2016)Regmi, Berthelot, Winkler, Mivelle, Proust,
  Bedu, Ozerov, Begou, Lumeau, Rigneault, \latin{et~al.} others]{regmi2016all}
Regmi,~R.; Berthelot,~J.; Winkler,~P.~M.; Mivelle,~M.; Proust,~J.; Bedu,~F.;
  Ozerov,~I.; Begou,~T.; Lumeau,~J.; Rigneault,~H., \latin{et~al.}
  All-dielectric silicon nanogap antennas to enhance the fluorescence of single
  molecules. \emph{Nano Lett.} \textbf{2016}, \emph{16}, 5143--5151\relax
\mciteBstWouldAddEndPuncttrue
\mciteSetBstMidEndSepPunct{\mcitedefaultmidpunct}
{\mcitedefaultendpunct}{\mcitedefaultseppunct}\relax
\EndOfBibitem
\bibitem[Rocco \latin{et~al.}(2020)Rocco, Lamprianidis, Miroshnichenko, and
  De~Angelis]{rocco2020giant}
Rocco,~D.; Lamprianidis,~A.; Miroshnichenko,~A.~E.; De~Angelis,~C. Giant
  electric and magnetic Purcell factor in dielectric oligomers. \emph{J. Opt.
  Soc. Am. B} \textbf{2020}, \emph{37}, 2738--2744\relax
\mciteBstWouldAddEndPuncttrue
\mciteSetBstMidEndSepPunct{\mcitedefaultmidpunct}
{\mcitedefaultendpunct}{\mcitedefaultseppunct}\relax
\EndOfBibitem
\bibitem[Yang \latin{et~al.}(2018)Yang, Zenin, and
  Bozhevolnyi]{yang2018anapole}
Yang,~Y.; Zenin,~V.~A.; Bozhevolnyi,~S.~I. Anapole-assisted strong field
  enhancement in individual all-dielectric nanostructures. \emph{ACS Photonics}
  \textbf{2018}, \emph{5}, 1960--1966\relax
\mciteBstWouldAddEndPuncttrue
\mciteSetBstMidEndSepPunct{\mcitedefaultmidpunct}
{\mcitedefaultendpunct}{\mcitedefaultseppunct}\relax
\EndOfBibitem
\bibitem[Mignuzzi \latin{et~al.}(2019)Mignuzzi, Vezzoli, Horsley, Barnes,
  Maier, and Sapienza]{mignuzzi2019nanoscale}
Mignuzzi,~S.; Vezzoli,~S.; Horsley,~S.~A.; Barnes,~W.~L.; Maier,~S.~A.;
  Sapienza,~R. Nanoscale design of the local density of optical states.
  \emph{Nano Lett.} \textbf{2019}, \emph{19}, 1613--1617\relax
\mciteBstWouldAddEndPuncttrue
\mciteSetBstMidEndSepPunct{\mcitedefaultmidpunct}
{\mcitedefaultendpunct}{\mcitedefaultseppunct}\relax
\EndOfBibitem
\bibitem[Albella \latin{et~al.}(2015)Albella, Shibanuma, and
  Maier]{albella2015switchable}
Albella,~P.; Shibanuma,~T.; Maier,~S.~A. Switchable directional scattering of
  electromagnetic radiation with subwavelength asymmetric silicon dimers.
  \emph{Sci. Rep.} \textbf{2015}, \emph{5}, 18322\relax
\mciteBstWouldAddEndPuncttrue
\mciteSetBstMidEndSepPunct{\mcitedefaultmidpunct}
{\mcitedefaultendpunct}{\mcitedefaultseppunct}\relax
\EndOfBibitem
\bibitem[Almeida \latin{et~al.}(2004)Almeida, Xu, Barrios, and
  Lipson]{almeida2004guiding}
Almeida,~V.~R.; Xu,~Q.; Barrios,~C.~A.; Lipson,~M. Guiding and confining light
  in void nanostructure. \emph{Opt. Lett.} \textbf{2004}, \emph{29},
  1209--1211\relax
\mciteBstWouldAddEndPuncttrue
\mciteSetBstMidEndSepPunct{\mcitedefaultmidpunct}
{\mcitedefaultendpunct}{\mcitedefaultseppunct}\relax
\EndOfBibitem
\bibitem[Galli \latin{et~al.}(2006)Galli, Gerace, Politi, Liscidini, Patrini,
  Andreani, Canino, Miritello, Savio, Irrera, \latin{et~al.}
  others]{galli2006direct}
Galli,~M.; Gerace,~D.; Politi,~A.; Liscidini,~M.; Patrini,~M.; Andreani,~L.~C.;
  Canino,~A.; Miritello,~M.; Savio,~R.~L.; Irrera,~A., \latin{et~al.}  Direct
  evidence of light confinement and emission enhancement in active
  silicon-on-insulator slot waveguides. \emph{Appl. Phys. Lett.} \textbf{2006},
  \emph{89}, 241114\relax
\mciteBstWouldAddEndPuncttrue
\mciteSetBstMidEndSepPunct{\mcitedefaultmidpunct}
{\mcitedefaultendpunct}{\mcitedefaultseppunct}\relax
\EndOfBibitem
\bibitem[Galli \latin{et~al.}(2006)Galli, Politi, Belotti, Gerace, Liscidini,
  Patrini, Andreani, Miritello, Irrera, Priolo, \latin{et~al.}
  others]{galli2006strong}
Galli,~M.; Politi,~A.; Belotti,~M.; Gerace,~D.; Liscidini,~M.; Patrini,~M.;
  Andreani,~L.; Miritello,~M.; Irrera,~A.; Priolo,~F., \latin{et~al.}  Strong
  enhancement of Er 3+ emission at room temperature in silicon-on-insulator
  photonic crystal waveguides. \emph{Appl. Phys. Lett.} \textbf{2006},
  \emph{88}, 251114\relax
\mciteBstWouldAddEndPuncttrue
\mciteSetBstMidEndSepPunct{\mcitedefaultmidpunct}
{\mcitedefaultendpunct}{\mcitedefaultseppunct}\relax
\EndOfBibitem
\bibitem[Jun \latin{et~al.}(2009)Jun, Briggs, Atwater, and
  Brongersma]{jun2009broadband}
Jun,~Y.~C.; Briggs,~R.~M.; Atwater,~H.~A.; Brongersma,~M.~L. Broadband
  enhancement of light emission in silicon slot waveguides. \emph{Opt. Express}
  \textbf{2009}, \emph{17}, 7479--7490\relax
\mciteBstWouldAddEndPuncttrue
\mciteSetBstMidEndSepPunct{\mcitedefaultmidpunct}
{\mcitedefaultendpunct}{\mcitedefaultseppunct}\relax
\EndOfBibitem
\bibitem[Kolchin \latin{et~al.}(2015)Kolchin, Pholchai, Mikkelsen, Oh, Ota,
  Islam, Yin, and Zhang]{kolchin2015high}
Kolchin,~P.; Pholchai,~N.; Mikkelsen,~M.~H.; Oh,~J.; Ota,~S.; Islam,~M.~S.;
  Yin,~X.; Zhang,~X. High Purcell factor due to coupling of a single emitter to
  a dielectric slot waveguide. \emph{Nano Lett.} \textbf{2015}, \emph{15},
  464--468\relax
\mciteBstWouldAddEndPuncttrue
\mciteSetBstMidEndSepPunct{\mcitedefaultmidpunct}
{\mcitedefaultendpunct}{\mcitedefaultseppunct}\relax
\EndOfBibitem
\bibitem[Sakib and Ryckman(2020)Sakib, and Ryckman]{sakib2020design}
Sakib,~N.; Ryckman,~J.~D. Design of ultra-small mode area all-dielectric
  waveguides exploiting the vectorial nature of light. \emph{Opt. Lett.}
  \textbf{2020}, \emph{45}, 4730--4733\relax
\mciteBstWouldAddEndPuncttrue
\mciteSetBstMidEndSepPunct{\mcitedefaultmidpunct}
{\mcitedefaultendpunct}{\mcitedefaultseppunct}\relax
\EndOfBibitem
\bibitem[Robinson \latin{et~al.}(2005)Robinson, Manolatou, Chen, and
  Lipson]{robinson2005ultrasmall}
Robinson,~J.~T.; Manolatou,~C.; Chen,~L.; Lipson,~M. Ultrasmall mode volumes in
  dielectric optical microcavities. \emph{Phys. Rev. Lett.} \textbf{2005},
  \emph{95}, 143901\relax
\mciteBstWouldAddEndPuncttrue
\mciteSetBstMidEndSepPunct{\mcitedefaultmidpunct}
{\mcitedefaultendpunct}{\mcitedefaultseppunct}\relax
\EndOfBibitem
\bibitem[Choi \latin{et~al.}(2017)Choi, Heuck, and Englund]{choi2017self}
Choi,~H.; Heuck,~M.; Englund,~D. Self-similar nanocavity design with ultrasmall
  mode volume for single-photon nonlinearities. \emph{Phys. Rev. Lett.}
  \textbf{2017}, \emph{118}, 223605\relax
\mciteBstWouldAddEndPuncttrue
\mciteSetBstMidEndSepPunct{\mcitedefaultmidpunct}
{\mcitedefaultendpunct}{\mcitedefaultseppunct}\relax
\EndOfBibitem
\bibitem[Hu \latin{et~al.}(2018)Hu, Khater, Salas-Montiel, Kratschmer,
  Engelmann, Green, and Weiss]{hu2018experimental}
Hu,~S.; Khater,~M.; Salas-Montiel,~R.; Kratschmer,~E.; Engelmann,~S.;
  Green,~W.~M.; Weiss,~S.~M. Experimental realization of deep-subwavelength
  confinement in dielectric optical resonators. \emph{Sci. Adv.} \textbf{2018},
  \emph{4}, eaat2355\relax
\mciteBstWouldAddEndPuncttrue
\mciteSetBstMidEndSepPunct{\mcitedefaultmidpunct}
{\mcitedefaultendpunct}{\mcitedefaultseppunct}\relax
\EndOfBibitem
\bibitem[Lee \latin{et~al.}(2010)Lee, Eom, Chang, Huh, Sung, and
  Shin]{lee2010silicon}
Lee,~S.; Eom,~S.~C.; Chang,~J.~S.; Huh,~C.; Sung,~G.~Y.; Shin,~J.~H. A silicon
  nitride microdisk resonator with a 40-nm-thin horizontal air slot. \emph{Opt.
  Express} \textbf{2010}, \emph{18}, 11209--11215\relax
\mciteBstWouldAddEndPuncttrue
\mciteSetBstMidEndSepPunct{\mcitedefaultmidpunct}
{\mcitedefaultendpunct}{\mcitedefaultseppunct}\relax
\EndOfBibitem
\bibitem[Sun \latin{et~al.}(2007)Sun, Dong, Feng, Hong, Michel, Lipson, and
  Kimerling]{sun2007horizontal}
Sun,~R.; Dong,~P.; Feng,~N.-n.; Hong,~C.-y.; Michel,~J.; Lipson,~M.;
  Kimerling,~L. Horizontal single and multiple slot waveguides: optical
  transmission at $\lambda$= 1550 nm. \emph{Opt. Express} \textbf{2007},
  \emph{15}, 17967--17972\relax
\mciteBstWouldAddEndPuncttrue
\mciteSetBstMidEndSepPunct{\mcitedefaultmidpunct}
{\mcitedefaultendpunct}{\mcitedefaultseppunct}\relax
\EndOfBibitem
\bibitem[Miyazaki and Kurokawa(2006)Miyazaki, and
  Kurokawa]{miyazaki2006squeezing}
Miyazaki,~H.~T.; Kurokawa,~Y. Squeezing visible light waves into a 3-nm-thick
  and 55-nm-long plasmon cavity. \emph{Phys. Rev. Lett.} \textbf{2006},
  \emph{96}, 097401\relax
\mciteBstWouldAddEndPuncttrue
\mciteSetBstMidEndSepPunct{\mcitedefaultmidpunct}
{\mcitedefaultendpunct}{\mcitedefaultseppunct}\relax
\EndOfBibitem
\bibitem[Ma \latin{et~al.}(2009)Ma, Zhang, and Van~Keuren]{ma2009analysis}
Ma,~C.; Zhang,~Q.; Van~Keuren,~E. Analysis of symmetric and asymmetric
  nanoscale slab slot waveguides. \emph{Opt. Commun.} \textbf{2009},
  \emph{282}, 324--328\relax
\mciteBstWouldAddEndPuncttrue
\mciteSetBstMidEndSepPunct{\mcitedefaultmidpunct}
{\mcitedefaultendpunct}{\mcitedefaultseppunct}\relax
\EndOfBibitem
\bibitem[Not()]{Note-1}
This resonance condition is valid for waveguide modes with a near-zero
  reflection phase delay at the waveguide-air facets along the propagation
  direction. Otherwise, one should include both propagation phase delay and
  reflection phase delay.\relax
\mciteBstWouldAddEndPunctfalse
\mciteSetBstMidEndSepPunct{\mcitedefaultmidpunct}
{}{\mcitedefaultseppunct}\relax
\EndOfBibitem
\bibitem[Sauvan \latin{et~al.}(2013)Sauvan, Hugonin, Maksymov, and
  Lalanne]{sauvan2013theory}
Sauvan,~C.; Hugonin,~J.-P.; Maksymov,~I.; Lalanne,~P. Theory of the spontaneous
  optical emission of nanosize photonic and plasmon resonators. \emph{Phys.
  Rev. Lett.} \textbf{2013}, \emph{110}, 237401\relax
\mciteBstWouldAddEndPuncttrue
\mciteSetBstMidEndSepPunct{\mcitedefaultmidpunct}
{\mcitedefaultendpunct}{\mcitedefaultseppunct}\relax
\EndOfBibitem
\bibitem[Lalanne \latin{et~al.}(2018)Lalanne, Yan, Vynck, Sauvan, and
  Hugonin]{lalanne2018light}
Lalanne,~P.; Yan,~W.; Vynck,~K.; Sauvan,~C.; Hugonin,~J.-P. Light interaction
  with photonic and plasmonic resonances. \emph{Laser Photonics Rev.}
  \textbf{2018}, \emph{12}, 1700113\relax
\mciteBstWouldAddEndPuncttrue
\mciteSetBstMidEndSepPunct{\mcitedefaultmidpunct}
{\mcitedefaultendpunct}{\mcitedefaultseppunct}\relax
\EndOfBibitem
\bibitem[Kristensen \latin{et~al.}(2012)Kristensen, Van~Vlack, and
  Hughes]{kristensen2012generalized}
Kristensen,~P.~T.; Van~Vlack,~C.; Hughes,~S. Generalized effective mode volume
  for leaky optical cavities. \emph{Opt. Lett.} \textbf{2012}, \emph{37},
  1649--1651\relax
\mciteBstWouldAddEndPuncttrue
\mciteSetBstMidEndSepPunct{\mcitedefaultmidpunct}
{\mcitedefaultendpunct}{\mcitedefaultseppunct}\relax
\EndOfBibitem
\bibitem[Cogn{\'e}e \latin{et~al.}(2019)Cogn{\'e}e, Yan, La~China, Balestri,
  Intonti, Gurioli, Koenderink, and Lalanne]{cognee2019mapping}
Cogn{\'e}e,~K.; Yan,~W.; La~China,~F.; Balestri,~D.; Intonti,~F.; Gurioli,~M.;
  Koenderink,~A.; Lalanne,~P. Mapping complex mode volumes with cavity
  perturbation theory. \emph{Optica} \textbf{2019}, \emph{6}, 269--273\relax
\mciteBstWouldAddEndPuncttrue
\mciteSetBstMidEndSepPunct{\mcitedefaultmidpunct}
{\mcitedefaultendpunct}{\mcitedefaultseppunct}\relax
\EndOfBibitem
\bibitem[Lalanne \latin{et~al.}(2019)Lalanne, Yan, Gras, Sauvan, Hugonin,
  Besbes, Dem{\'e}sy, Truong, Gralak, Zolla, \latin{et~al.}
  others]{lalanne2019quasinormal}
Lalanne,~P.; Yan,~W.; Gras,~A.; Sauvan,~C.; Hugonin,~J.-P.; Besbes,~M.;
  Dem{\'e}sy,~G.; Truong,~M.; Gralak,~B.; Zolla,~F., \latin{et~al.}
  Quasinormal mode solvers for resonators with dispersive materials. \emph{J.
  Opt. Soc. Am. A} \textbf{2019}, \emph{36}, 686--704\relax
\mciteBstWouldAddEndPuncttrue
\mciteSetBstMidEndSepPunct{\mcitedefaultmidpunct}
{\mcitedefaultendpunct}{\mcitedefaultseppunct}\relax
\EndOfBibitem
\bibitem[Barnes \latin{et~al.}(2020)Barnes, Horsley, and
  Vos]{barnes2020classical}
Barnes,~W.~L.; Horsley,~S.~A.; Vos,~W.~L. Classical antennae, quantum emitters,
  and densities of optical states. \emph{J. Opt.} \textbf{2020}, \relax
\mciteBstWouldAddEndPunctfalse
\mciteSetBstMidEndSepPunct{\mcitedefaultmidpunct}
{}{\mcitedefaultseppunct}\relax
\EndOfBibitem
\bibitem[Krasnok \latin{et~al.}(2015)Krasnok, Slobozhanyuk, Simovski,
  Tretyakov, Poddubny, Miroshnichenko, Kivshar, and Belov]{krasnok2015antenna}
Krasnok,~A.~E.; Slobozhanyuk,~A.~P.; Simovski,~C.~R.; Tretyakov,~S.~A.;
  Poddubny,~A.~N.; Miroshnichenko,~A.~E.; Kivshar,~Y.~S.; Belov,~P.~A. An
  antenna model for the Purcell effect. \emph{Sci. Rep.} \textbf{2015},
  \emph{5}, 12956\relax
\mciteBstWouldAddEndPuncttrue
\mciteSetBstMidEndSepPunct{\mcitedefaultmidpunct}
{\mcitedefaultendpunct}{\mcitedefaultseppunct}\relax
\EndOfBibitem
\bibitem[Denning \latin{et~al.}(2018)Denning, Iles-Smith, Osterkryger,
  Gregersen, and Mork]{denning2018cavity}
Denning,~E.~V.; Iles-Smith,~J.; Osterkryger,~A.~D.; Gregersen,~N.; Mork,~J.
  Cavity-waveguide interplay in optical resonators and its role in optimal
  single-photon sources. \emph{Phys. Rev. B} \textbf{2018}, \emph{98},
  121306\relax
\mciteBstWouldAddEndPuncttrue
\mciteSetBstMidEndSepPunct{\mcitedefaultmidpunct}
{\mcitedefaultendpunct}{\mcitedefaultseppunct}\relax
\EndOfBibitem
\bibitem[Not()]{Note-2}
It should be noted that the emitter is at the position of peak field intensity
  for both resonant modes of the structure. Hence, the dipole mostly decays
  through coupling to the antenna modes and negligibly through direct coupling
  to free-space modes.\relax
\mciteBstWouldAddEndPunctfalse
\mciteSetBstMidEndSepPunct{\mcitedefaultmidpunct}
{}{\mcitedefaultseppunct}\relax
\EndOfBibitem
\bibitem[Warne \latin{et~al.}(2012)Warne, Basilio, Langston, Johnson, and
  Sinclair]{warne2012perturbation}
Warne,~L.~K.; Basilio,~L.~I.; Langston,~W.~L.; Johnson,~W.~A.; Sinclair,~M.~B.
  Perturbation theory in the design of degenerate rectangular dielectric
  resonators. \emph{Prog. Electromagn. Res.} \textbf{2012}, \emph{44},
  1--29\relax
\mciteBstWouldAddEndPuncttrue
\mciteSetBstMidEndSepPunct{\mcitedefaultmidpunct}
{\mcitedefaultendpunct}{\mcitedefaultseppunct}\relax
\EndOfBibitem
\bibitem[Staude \latin{et~al.}(2013)Staude, Miroshnichenko, Decker, Fofang,
  Liu, Gonzales, Dominguez, Luk, Neshev, Brener, \latin{et~al.}
  others]{staude2013tailoring}
Staude,~I.; Miroshnichenko,~A.~E.; Decker,~M.; Fofang,~N.~T.; Liu,~S.;
  Gonzales,~E.; Dominguez,~J.; Luk,~T.~S.; Neshev,~D.~N.; Brener,~I.,
  \latin{et~al.}  Tailoring directional scattering through magnetic and
  electric resonances in subwavelength silicon nanodisks. \emph{ACS Nano}
  \textbf{2013}, \emph{7}, 7824--7832\relax
\mciteBstWouldAddEndPuncttrue
\mciteSetBstMidEndSepPunct{\mcitedefaultmidpunct}
{\mcitedefaultendpunct}{\mcitedefaultseppunct}\relax
\EndOfBibitem
\bibitem[Liu and Kivshar(2018)Liu, and Kivshar]{liu2018generalized}
Liu,~W.; Kivshar,~Y.~S. Generalized Kerker effects in nanophotonics and
  meta-optics. \emph{Opt. Express} \textbf{2018}, \emph{26}, 13085--13105\relax
\mciteBstWouldAddEndPuncttrue
\mciteSetBstMidEndSepPunct{\mcitedefaultmidpunct}
{\mcitedefaultendpunct}{\mcitedefaultseppunct}\relax
\EndOfBibitem
\bibitem[Balanis(2016)]{balanis2016antenna}
Balanis,~C.~A. \emph{Antenna theory: analysis and design}; John wiley \& sons,
  2016\relax
\mciteBstWouldAddEndPuncttrue
\mciteSetBstMidEndSepPunct{\mcitedefaultmidpunct}
{\mcitedefaultendpunct}{\mcitedefaultseppunct}\relax
\EndOfBibitem
\bibitem[Pursula \latin{et~al.}(2007)Pursula, Hirvonen, Jaakkola, and
  Varpula]{pursula2007antenna}
Pursula,~P.; Hirvonen,~M.; Jaakkola,~K.; Varpula,~T. Antenna effective aperture
  measurement with backscattering modulation. \emph{IEEE Trans. Antennas
  Propag.} \textbf{2007}, \emph{55}, 2836--2843\relax
\mciteBstWouldAddEndPuncttrue
\mciteSetBstMidEndSepPunct{\mcitedefaultmidpunct}
{\mcitedefaultendpunct}{\mcitedefaultseppunct}\relax
\EndOfBibitem
\bibitem[Not()]{Note-3}
The spectral directivity values of PGAs in the receiving configuration are
  similar to those of transmission. However, both values differ from each other
  when the emitter is weakly coupled to the resonant modes of the structure.
  For an accurate characterization of directivity during reception, the
  resonant modes should be excited by a plane wave propagating towards the
  structure in the desired direction.\relax
\mciteBstWouldAddEndPunctfalse
\mciteSetBstMidEndSepPunct{\mcitedefaultmidpunct}
{}{\mcitedefaultseppunct}\relax
\EndOfBibitem
\end{mcitethebibliography}

\begin{suppinfo}

\beginsupplement

\begin{figure}[b]
\centering 
\includegraphics[width=0.8\textwidth]{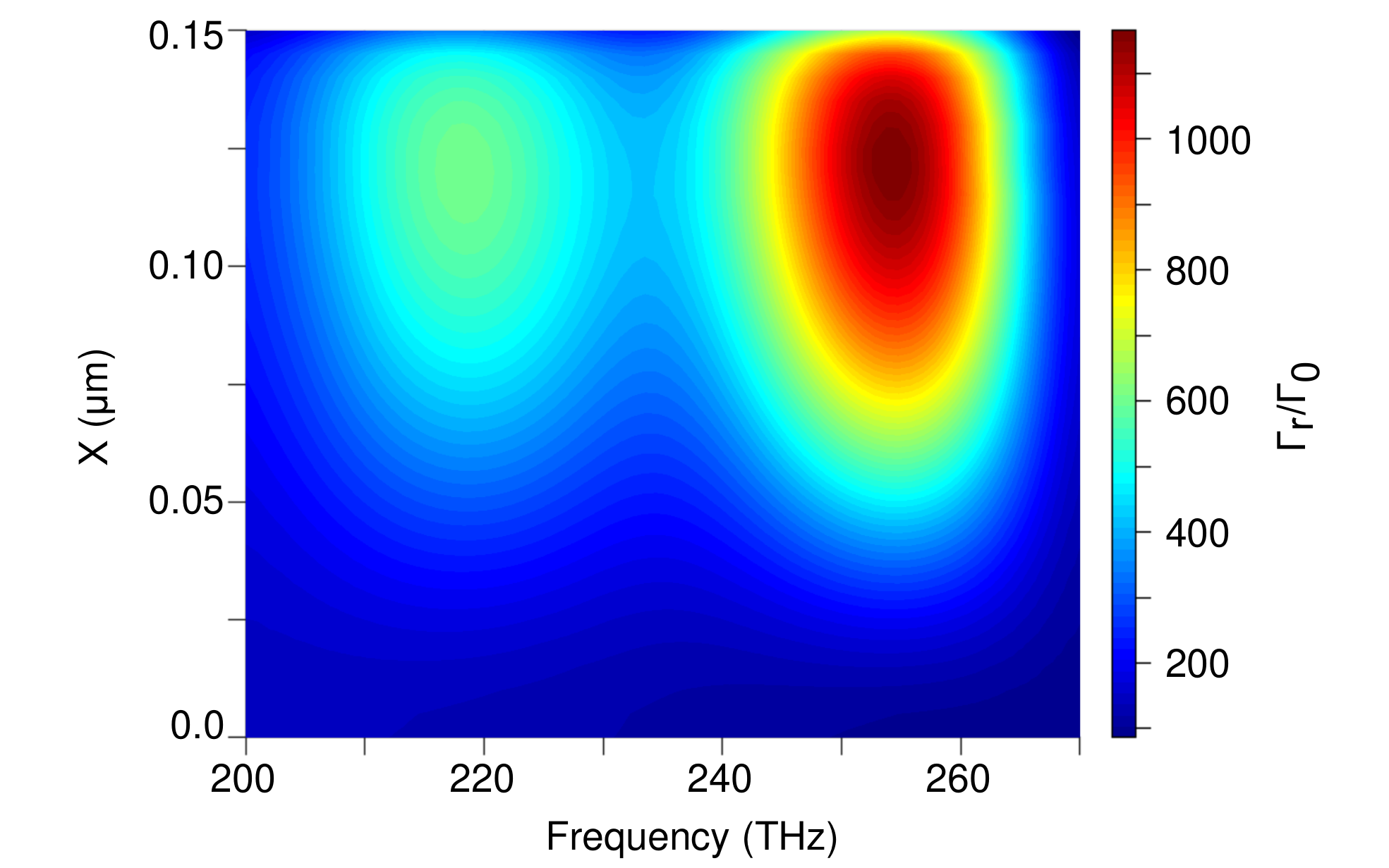}
\caption{SER enhancement factors versus frequency for the variation in the lateral position of the emitter (along $x$, from the center to the edge) embedded within the gap of the asymmetric PGA shown in Fig.~\ref{fig:StructurePurcellFbr}(a) of main text.} \label{fig:SER_lateral}
\end{figure}

\begin{figure}[b]
\centering 
\includegraphics[width=0.8\textwidth]{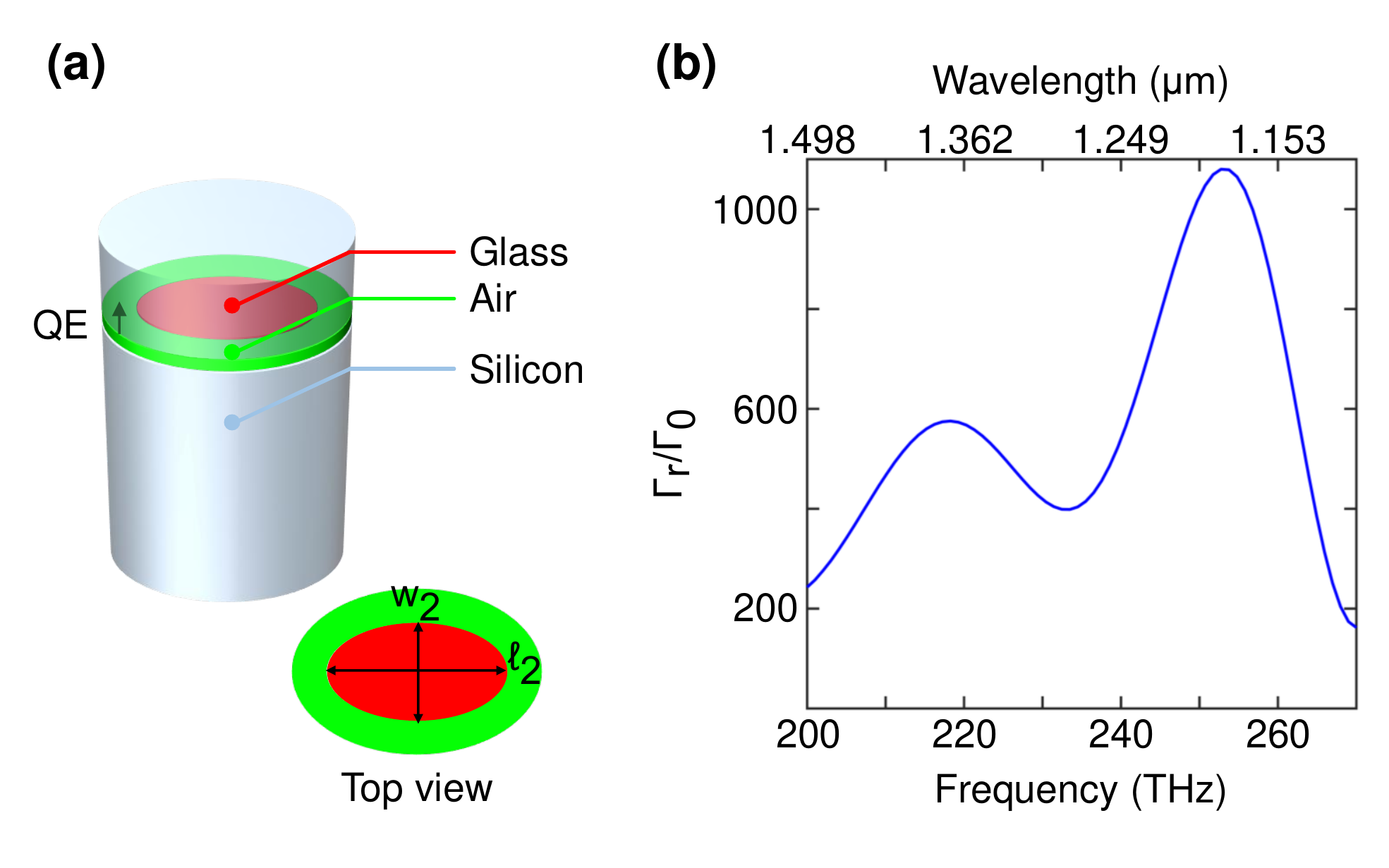}
\caption{Design and emission properties of asymmetric PGA with air gap and glass spacer. (a) Perspective view and top view of PGA with a glass spacer in the air gap layer. The design parameters of the PGA are the same as in Fig.~\ref{fig:StructurePurcellFbr}(a). The glass spacer has a thickness of 2 nm, and $w_2$ and $l_2$ of 140 nm and 200 nm, respectively. (b) SER enhancement factors versus frequencies for the asymmetric PGA shown in (a) for the same position (orientation) of the quantum emitter (QE) as in Fig.~\ref{fig:StructurePurcellFbr}(a) of main text.} \label{fig:SER_fabrication}
\end{figure}

\begin{figure}[b]
\centering 
\includegraphics[width=0.8\textwidth]{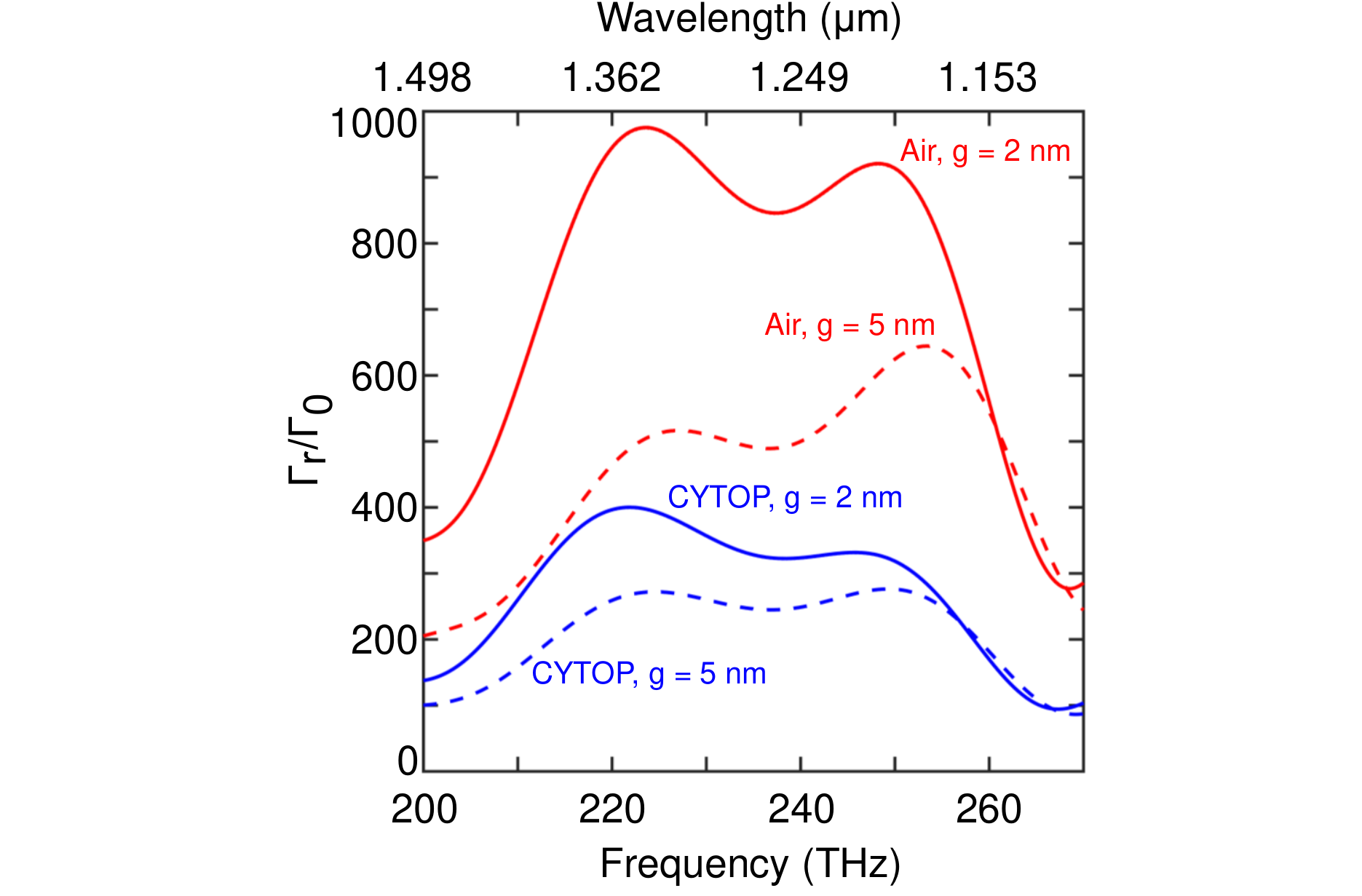}
\caption{SER enhancement factors versus frequency for PGAs on glass substrate ($n=1.50$). The design parameters and the position of the dipole emitter are same as that of Fig.~\ref{fig:improvedbandwidthdirectionality}(b) of main text.} \label{fig:SER_substrate}
\end{figure}

\begin{figure}[b]
\centering 
\includegraphics[width=0.8\textwidth]{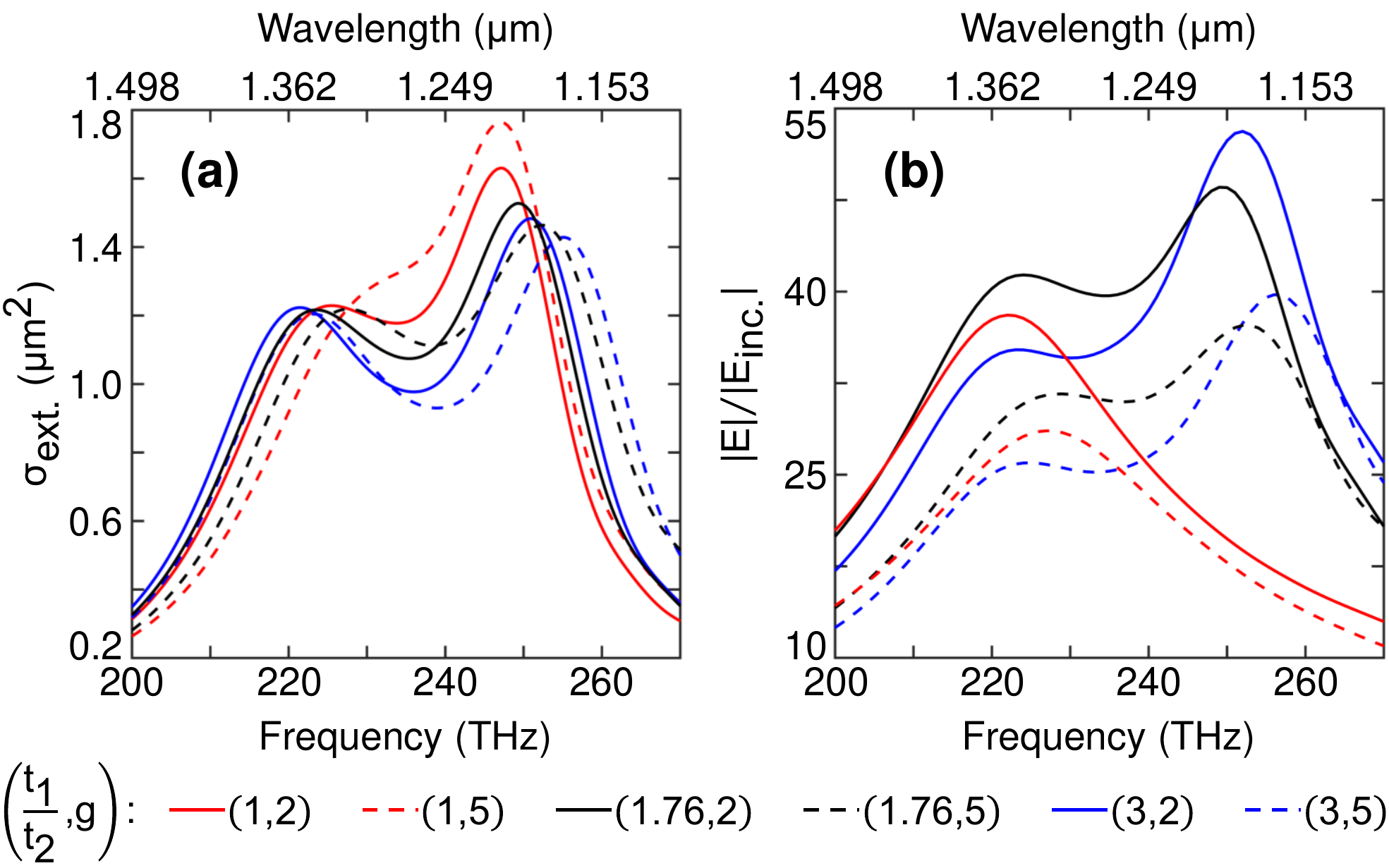}
\caption{Extinction cross-sections and electric field enhancement factor of PGAs with air gap. (a) Extinction cross-section ($\sigma_{\text{ext.}}$) versus frequency, and (b) field enhancement factor versus frequency for the elliptical PGAs with air gaps, where the gap position and thickness are varied. The combined thickness of silicon ($t_1+t_2=580$~nm), $w_1$ and $\ell_1$ are same as for the PGA in Fig.~\ref{fig:StructurePurcellFbr}(a) of main text.} \label{fig:CrosssecFieldenahnceWithair}
\end{figure}

\end{suppinfo}

\end{document}